\documentclass[prl,aps,twocolumn,amsmath,amssymb,floatfix,superscriptaddress,preprintnumbers]{revtex4-2}
\usepackage{graphicx}% Include figure files
\usepackage{dcolumn}% Align table columns on decimal point
\usepackage{bm}% bold math
\usepackage{color}

\font\scripti=cmmi7
\font\scriptscripti=cmmi5
\def\sib#1{\setbox0 = \hbox{\scripti #1}
  \kern-.02em\copy0\kern-\wd0
  \kern.04em\box0} % script italic bold 
\def\ssib#1{\setbox0 = \hbox{\scriptscripti #1}
  \kern-.02em\copy0\kern-\wd0
  \kern.04em\box0} % scriptscript italic bold
\font\tenib=cmmib10 % italic bold for math
\skewchar\tenib='177 \skewchar\tenib='177 \skewchar\tenib='177
\def\pbold#1{\setbox0 = \hbox{$ #1 $}
  \kern-.022em\copy0\kern-\wd0
  \kern.011em\copy0\kern-\wd0
  \kern.011em\copy0\kern-\wd0
  \kern.011em\copy0\kern-\wd0
  \kern.011em\box0} % poorman's bold
\def\up{\uparrow}
\def\dwn{\downarrow}

\def\lesssim{\ \raise.3ex\hbox{$<$}\kern-0.8em\lower.7ex\hbox{$\sim$}\ }
\def\gesim{\ \raise.3ex\hbox{$>$}\kern-0.8em\lower.7ex\hbox{$\sim$}\ }

\newcommand{\beginsupplement}{%
        \setcounter{table}{0}
        \renewcommand{\thetable}{S\arabic{table}}%
        \setcounter{figure}{0}
        \renewcommand{\thefigure}{S\arabic{figure}}%
      \setcounter{equation}{0}
        \renewcommand{\theequation}{S.\arabic{equation}}%

     }

\begin{document}
\preprint{NITEP 170}

\title{Polaronic Proton and Diproton Clustering in Neutron-Rich Matter}
\author{Hiroyuki Tajima}
\affiliation{Department of Physics, Graduate School of Science, The University of Tokyo, Tokyo 113-0033, Japan}
\affiliation{RIKEN Nishina Center, Wako 351-0198, Japan}
\author{Hajime Moriya}
\affiliation{Department of Physics, Hokkaido University, Sapporo 060-0810,
  Japan}
\author{Wataru Horiuchi}
\affiliation{Department of Physics, Osaka Metropolitan University, Osaka 558-8585, Japan}
\affiliation{Nambu Yoichiro Institute of Theoretical and Experimental Physics (NITEP), Osaka Metropolitan University, Osaka 558-8585, Japan}
\affiliation{RIKEN Nishina Center, Wako 351-0198, Japan}
\affiliation{Department of Physics, Hokkaido University, Sapporo 060-0810,
  Japan}
\author{Eiji Nakano}
\affiliation{Department of Mathematics and Physics, Kochi University, Kochi 780-8520, Japan}
\author{Kei Iida}
\affiliation{Department of Mathematics and Physics, Kochi University, Kochi 780-8520, Japan}

\date{\today}
\begin{abstract}
We show that strong spin-triplet neutron-proton interaction causes polaronic protons to occur in neutron matter at subnuclear densities and nonzero temperature.
As the neutron density increases, proton spectra exhibit a smooth crossover from a bare impurity to a repulsive polaron branch; this branch coexists with an attractive polaron branch.  With the neutron density increased further, the attractive polarons become stable with respect to deuteron formation.  For two adjacent protons, we find that the polaron effects and the neutron-mediated attraction are sufficient to induce a bound diproton, which leads possibly to diproton formation in the surface region of neutron-rich nuclei in laboratories as well as in neutron stars. 
\end{abstract}
%\pacs{03.75.Ss, 03.75.-b, 03.70.+k}
\maketitle
%%%%%%%%%%%%%%%%%%%%%%%%%%%%%%%%%%%%%%%%%%%%%%%%%%%%%%%%%%%%%%%%%%%%%%%%%%%%%
%\par
%\section{Introduction}
%\section{Introduction}
\noindent{\it Introduction}---
Just after the Bardeen-Cooper-Schrieffer (BCS) theory was developed for superconductivity in metals~\cite{PhysRev.108.1175}, it was immediately applied to describe nuclei and hadrons; superfluidity in nuclei was adovocated by Bohr, Mottelson, and Pines~\cite{PhysRev.110.936}, while Nambu and Jona-Lasinio~\cite{PhysRev.122.345,PhysRev.124.246} advanced a theory of nucleons and mesons in terms of chiral symmetry breaking.  These are typical examples showing that quantum many-body states underlie subatomic particles even in vacuum.  If such particles are embedded in many nucleon environments as encountered in neutron stars, it would be still more interesting~\cite{PhysRevC.81.015803}.  Sure enough, nuclei and hadrons as {\em impurities} could be significantly modified from the corresponding vacuum states~\cite{RevModPhys.82.2949}.  Here we address 
how such modifications affect the pairing properties of the impurities.

%Introduction for polarons in cold atomic systems
Understanding quantum
properties of an impurity immersed in medium dated back to the notion of a polaron, which was proposed by Landau and Pekkar~\cite{landau1933electron,landau1948effective} even before the advent of the BCS theory. 
While it was originally used for the description of electrons in ionic lattices, which was later analyzed for a weakly coupled electron-phonon system by Lee, Low, and Pines (LLP)~\cite{PhysRev.90.297}, recently, trapped cold atoms act as quantum simulators that provide an ideal platform to investigate the polaron problem in various settings.
In particular, minority atoms immersed in a Fermi (Bose) gas are referred to as Fermi (Bose) polarons~\cite{massignan2014polarons,atoms10020055}.
Various properties of polarons
have been measured experimentally~\cite{PhysRevLett.102.230402,PhysRevLett.103.170402,nascimbene2010exploring,PhysRevLett.108.235302,PhysRevLett.117.055301,PhysRevLett.117.055302,PhysRevLett.122.093401,yan2020bose}.
When the scattering length is positive between minority and majority atoms, moreover, an excited polaronic state called repulsive polarons~\cite{koschorreck2012attractive,PhysRevLett.118.083602,PhysRevLett.125.133401} is known to appear in addition to the ground-state attractive polarons.
At a certain coupling strength, polaron-molecule transition has also been found experimentally~\cite{PhysRevX.10.041019}.
The medium-induced interaction between two minority atoms, which depends on the statistics of the two species,
has also been investigated experimentally~\cite{desalvo2019observation,PhysRevLett.124.163401,PhysRevA.103.053314}.

\begin{figure}[t]
    \centering
    \includegraphics[width=6cm]{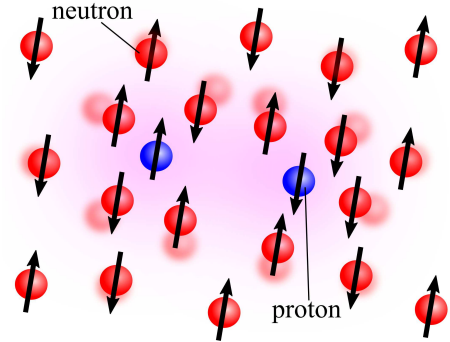}
    \caption{Schematics for two adjacent protons immersed in neutron matter.  Each proton is dressed with a cloud of neutron excitations due to the strong neutron-proton interaction and forms a Fermi polaron.
    Such polaronic protons may form a two-body bound state (diproton) in neutron matter although in vacuum the proton-proton interaction is not strong enough to form a diproton.}
    \label{fig:1}
\end{figure}

%Introduction for polarons in nuclear systems
Given the similarity of ultracold Fermi gases with dilute neutron matter~\cite{horikoshi2019cold}, it is natural to consider that
the concept of polarons 
can be applied to nuclear systems. 
In this context, an alpha particle immersed in dilute neutron matter has been studied by some of the present authors~\cite{PhysRevC.102.055802} within a theoretical framework developed in cold-atom physics~\cite{PhysRevA.74.063628}.
Remarkably, unbound clusters (e.g., $^8$Be,  the Hoyle state~\cite{hoyle1954resonances}) could turn into bound states due to the medium effect~\cite{PhysRevC.104.065801,PhysRevC.106.045807}.
For nucleon impurities, 
a polaronic neutron in spin-polarized neutron matter has been considered as a benchmark for theoretical studies~\cite{PhysRevC.89.041301,PhysRevC.103.L052801}.
Moreover, a proton impurity
immersed in neutron matter (see also Fig.~\ref{fig:1}),
which would be relevant to neutron star matter with small proton fraction~\cite{RevModPhys.89.015007},
has been studied within the LLP theory for a weakly coupled proton-phonon system~\cite{PhysRevC.47.1077}.
However, such a mixture involves the strong spin-triplet neutron-proton interaction responsible for deuteron formation~\cite{PhysRevLett.83.1926,ROPKE199953,PhysRevC.81.034007,PhysRevC.82.024911,PhysRevC.96.034327,PhysRevC.99.014310}.
It is instructive to recall from microscopic analyses \cite{li2018nucleon}
that for neutron star matter around or above normal nuclear density, the proton effective mass in Landau's Fermi liquid theory is smaller than the bare mass and increases
with decreasing density.  Although we are interested in the dilute limit of protons where protons are no longer Fermi degenerate, there may be
connections between the polaronic and Fermi-liquid properties.
Anyway, it is important to incorporate such a strong neutron-proton interaction to understand how protons behave under an
extremely neutron-rich environment. 

In this Letter, 
we consider the fate of a proton immersed in neutron matter at subnuclear densities and nonzero temperatures by building 
the spin-triplet neutron-proton interaction, leading to the only known two-nucleon bound state in vacuum, namely, a deuteron,
into a field-theoretical formalism developed for atomic Fermi polarons, that is, many-body $T$-matrix approach (TMA)~\cite{PhysRevA.78.031602,PhysRevLett.98.180402,PhysRevA.98.013626,atoms9010018,PhysRevA.105.043303}.
Quasiparticle properties of each proton impurity, together with induced interaction between two adjacent impurities,
may lead to a bound diproton, as shown in Fig.~\ref{fig:1}.
This is similar to the case of two adjacent alpha particles in dilute neutron matter~\cite{PhysRevC.104.065801},
in the sense that a two-impurity state, unbound in vacuum, becomes bound in neutron matter. 
These impurities may have possible relevance to the structure and evolution of the deepest region of neutron star crusts~\cite{pethick1995matter},
as well as the clustering in the surface region of neutron-rich nuclei~\cite{tanaka2021formation}.

\noindent{\it Formalism}---
We use units in which $\hbar=k_{\rm B}=1$ and set the system volume to unity.
We begin with the Hamiltonian for a proton-neutron mixture that interacts via a spin-triplet, isoscalar 
potential:
\begin{align}
\label{eq:H}
    H&=\sum_{\sigma=\pm 1/2}\sum_{\tau=\pm 1/2}\sum_{\bm{k}}\xi_{\bm{k},\tau} c_{\bm{k},\sigma,\tau}^\dag c_{\bm{k},\sigma,\tau} \cr
    &+ \frac{1}{2}\sum_{S_z=-1}^{S_z=+1}   T^{\dagger}_{S_z} (\bm{k},\bm{q}) V_{\rm t}(\bm{k},\bm{k}')  T_{S_z} (\bm{k}',\bm{q}),
\end{align}
where $\xi_{\bm{k},\tau}=k^2/2M-\mu_\tau$ and
$c_{\bm{k},\sigma,\tau}^{(\dag)}$ are respectively the kinetic energy with the nucleon mass $M=939$ MeV and the annihilation (creation) operator of a nucleon with momentum $\bm{k}$, spin $\sigma=\pm1/2 $, and isospin $\tau=\pm 1/2$. $\mu_{\tau}$ is the nucleon chemical potential. 
The interaction term involves the spin-triplet pair operator with $z$-component of the total spin $S_z=0,\pm 1$, relative momentum $\bm{k}$, and center-of-mass momentum $\bm{q}$ as given by
\begin{align}
    T_{S_z} (\bm{k},\bm{q}) &= \sum_{\sigma, \sigma'} \sum_{\tau,\tau'} 
\Bigl\langle \frac{1}{2} \frac{1}{2} \sigma \sigma' \Bigr|\Bigl.1 S_z \Bigr\rangle 
\Bigr\langle \frac{1}{2} \frac{1}{2} \tau\tau' \Bigr|\Bigl. 00 \Bigr\rangle\cr 
&\quad \times  c_{\bm{-k+q}/2,\sigma,\tau}  c_{\bm{k+q}/2,\sigma',\tau'}.
\end{align}
In what follows, for convenience the isospin $\tau$ is explicitly expressed as $\tau=+1/2\equiv {\rm n}$ (neutron) and $\tau=-1/2\equiv {\rm p}$ (proton),  and we ignore isospin symmetry breaking contributions including Coulomb interactions.
The explicit form of $H$ can be found in the Supplement~\cite{Supplement}.

For the spin-triplet (isospin-singlet) channel, we employ the Yamaguchi-type separable interaction
    $V_{\rm t}(\bm{k},\bm{k}')=-\gamma_k\gamma_{k'}$,
where $\gamma_k=u_{\rm t}/(k^2+\Lambda_{\rm t}^2)$ is the form factor~\cite{YamaguchiPhysRev.95.1628}.
The parameters $u_{\rm t}$ and $\Lambda_{\rm t}$ are related to the spin-triplet scattering length $a_{\rm t}$ and effective range 
$r_{\rm t}$ as 
    $u_{\rm t}=\Lambda_{\rm t}^2\sqrt{\frac{8\pi}{M}\frac{1}{\Lambda_{\rm t}-2/a_{\rm t}}}$,
    and
    $\Lambda_{\rm t}=\frac{3+\sqrt{9-16r_{\rm t}/a_{\rm t}}}{2r_{\rm t}}$~\cite{tajima2019superfluid}.
For the empirical values $a_{\rm t}=5.42$ fm and $r_{\rm t}=1.76$ fm~\cite{PhysRevC.51.38},
the resultant $u_{\rm t}$ and $\Lambda_{\rm t}$ lead to the deuteron binding energy $E_{\rm d}=2.2$~MeV, which is consistent with
the empirical one, and reproduce the empirical $^3S_1$ phase shift well up to $k\simeq 1$ fm$^{-1}$~\cite{tajima2019superfluid}.

Effects of the strong spin-triplet neutron-proton interaction are incorporated via
the in-medium neutron-proton $T$-matrix
\begin{align}
    \Gamma_{\sigma\sigma'}(\bm{k},\bm{k}';\bm{q},i\nu_\ell)=-\frac{\gamma_k\gamma_{k'}}{1-\Pi(\bm{q},i\nu_\ell)}\frac{\delta_{\sigma,\sigma'}+1}{2},
\end{align}
with the neutron-proton propagator
\begin{align}
    \Pi(\bm{q},i\nu_\ell)=-\sum_{\bm{k}}\frac{\gamma_k^2[1-f(\xi_{\bm{k}+\bm{q}/2,{\rm n}})]}{i\nu_n-\xi_{\bm{k}+\bm{q}/2,{\rm n}}-\xi_{-\bm{k}+\bm{q}/2,{\rm p}}}.
\end{align}
For more details, see also the Supplement~\cite{Supplement}.
In this work, TMA, which reproduces various Fermi-polaron properties in cold atomic systems, is used to describe the polaronic state of a proton in neutron matter.
The proton self-energy is given by~\cite{Supplement}
\begin{align}
\label{eq:sigma}
    &\Sigma_{\rm p\sigma}(\bm{k},\omega)
= 
    \sum_{\bm{q}}\sum_{\sigma'}f(\xi_{\bm{q}-\bm{k},{\rm n}})\cr
    &
    \quad
    \times\Gamma_{\sigma\sigma'}(\bm{q}/2-\bm{k},\bm{q}/2-\bm{k};\bm{q},\omega_+-\mu_{\rm p}+\xi_{\bm{q}-\bm{k},{\rm n}}),
\end{align}
with $\omega_+=\omega+i\delta$ and an infinitesimally small number $\delta$ (practically, $2M\delta=10^{-2}$ fm$^{-2}$ is taken in the numerical calculation), where $f(\xi_{\bm{q}-\bm{k},{\rm n}})=\left(e^{\xi_{\bm{q}-\bm{k},{\rm n}}/T}+1\right)^{-1}$ is the neutron Fermi-Dirac distribution function with the temperature $T$.
The proton Green's function reads $G_{{\rm p}\sigma}(\bm{k},\omega)=\left[\omega_+-\frac{k^2}{2M}-\Sigma_{\rm p\sigma}(\bm{k},\omega)\right]^{-1}$.
Furthermore, $G_{{\rm p}\sigma}(\bm{k},\omega)$ can be approximately expressed as
$G_{{\rm p}\sigma}(\bm{k},\omega)\simeq Z\left[\omega_+-\frac{k^2}{2M_{\rm eff}}-E_{\rm P}+i\Gamma_{\rm P}/2\right]^{-1}$,
where $Z$, $M_{\rm eff}$, $E_{\rm P}$, and $\Gamma_{\rm P}$ are the polaron residue, the effective mass, the polaron energy, and the decay rate, respectively.

In this Letter, we take the small proton-fraction limit (i.e., $\rho_{\rm p}/\rho_{\rm n}\rightarrow 0$ where $\rho_{\rm p}$ and $\rho_{\rm n}$ are the proton and neutron densities, respectively) by setting $\mu_{\rm p}\rightarrow-\infty$ at finite temperature.  In this limit, we can safely neglect backaction from protons on uniformity of the remaining neutron matter~\cite{atoms9010018}.  This matter is assumed to be a spin-balanced ideal gas in such a way as to be consistent with Eq.\ (\ref{eq:H}).  $\mu_{\rm n}$ can then be determined by solving the number equation $\rho_{\rm n}=2\sum_{\bm{k}}f(\xi_{\bm{k},{\rm n}})$.  Note that in a more realistic situation, neutron matter is a superfluid as long as the temperature is below the critical temperature $T_{\rm c}$, which is typically $\sim 1$ MeV~\cite{ramanan2021pairing}. 
We can nevertheless assume that neutron superfluid properties would make only a little difference in the polaronic properties~\cite{PhysRevA.105.023317}.

\noindent{\it Polaronic proton}---
\begin{figure}[t]
    \centering
    \includegraphics[width=8cm]{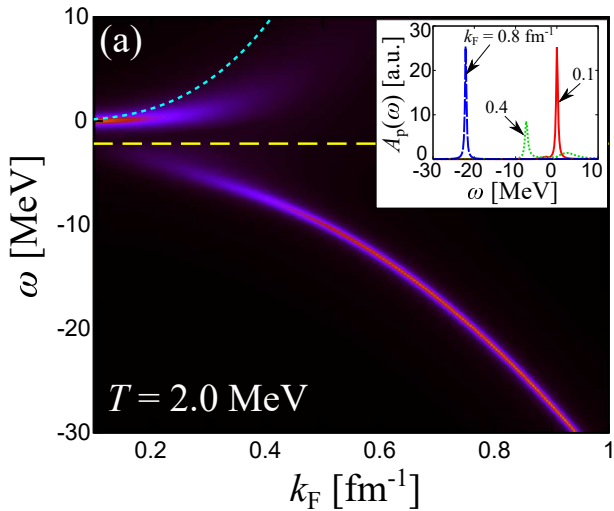}
    \caption{Polaronic proton spectral weight $A_{\rm p}(\omega)$ at $T=2.0$~MeV, plotted as a function of the neutron Fermi momentum $k_{\rm F}$.
 The color brightness indicates the intensity of $A_{\rm p}(\omega)$ in an arbitrary unit. The dotted line shows the Hartree shift, while the dashed line is the deuteron energy $-E_{\rm d}=-2.2$~MeV.
 The inset shows $A_{\rm p}(\omega)$ at $k_{\rm F}=0.1$~fm$^{-1}$, $0.4$~fm$^{-1}$, and $0.8$~fm$^{-1}$. 
 }
    \label{fig:2}
\end{figure}
Figure~\ref{fig:2} exhibits the proton spectral weight $A_{{\rm p}}(\omega)=-\frac{1}{\pi}{\rm Im}G_{{\rm p}\sigma}(\bm{k}=\bm{0},\omega)$ 
as a function of the neutron Fermi momentum $k_{\rm F}=(3\pi^2\rho_{\rm n})^{1/3}$.  The temperature $T$ is fixed at $2$~MeV (note that the result obtained at $T=0.1$ MeV in the Supplement~\cite{Supplement} is essentially the same). Then,
the dimensionless temperature $T/T_{\rm F}$ with the neutron Fermi temperature $T_{\rm F}=\frac{k_{\rm F}^2}{2M_{\rm F}}$ changes from $T/T_{\rm F}\sim 10$ to $T/T_{\rm F}\sim 0.1$ with increasing $k_{\rm F}$.
In the dilute regime ($k_{\rm F}\simeq0.1$ fm$^{-1}$), neutrons behave like a classical Boltzmann gas ($T\gg T_{\rm F}$) , while in
the high density regime ($k_{\rm F}\gtrsim0.3 \ {\rm fm}^{-1}$), they behave like a quantum degenerate gas ($T_{\rm F}\gtrsim T$).  In between, 
a crossover from bare impurities to repulsive polarons can be found for protons when $k_{\rm F}$ increases.  In fact,
in the classical regime ($T\gg T_{\rm F}$), the proton energy is close to zero, but as
the density approaches the quantum degenerate regime ($T_{\rm F}\simeq T$), 
the bare-impurity branch starts to follow the Hartree shift $\Sigma_{\rm H}=\frac{6\pi a_{\rm t}}{M}\rho_{\rm n}$ and also 
to be broadened, indicating a crossover towards the repulsive polaron branch with finite decay width.
Simultaneously, another broadened low-energy branch, that is, attractive polaron branch, appears in such a regime. 
Incidentally, a similar kind of coexistence of attractive and repulsive polarons is predicted from atomic Fermi polarons in the strong-coupling regime~\cite{PhysRevA.99.063606,PhysRevA.102.023304,PhysRevA.105.043303}.
For comparison, the deuteron energy $-E_{\rm d}=-2.2$ MeV is also plotted in Fig.~\ref{fig:2}.
At relatively high neutron densities where $E_{\rm F}\gg E_{\rm d}$, the attractive polaron is stabilized as evident from a sharp peak at low energy. In this situation, since $E_{\rm F}$ is larger than the neutron separation energy of the deuteron, deuteron-like molecules would no longer be bound.   
A possible alpha cluster state, if any in such a high-density regime, would tend to melt into free nucleons  
at low temperatures of interest here~\cite{PhysRevLett.80.3177,PhysRevC.82.034322}.

In the low-temperature limit, the polaron energy $E_{\rm P}$ defined as the peak position of $A_{\rm p}(\omega)$ in Fig.~\ref{fig:2} (see also Supplement~\cite{Supplement}) is associated with the nuclear equation of state (EOS) via the Landau-Pomeranchuk form of the total energy per nucleon $E/A=E_{\rm PNM}/A+E_{\rm P}\rho_{\rm p}/\rho_{\rm n}+O(\rho_{\rm p}^2/\rho_{\rm n}^2)$, where $A$ is the total nucleon number, and $E_{\rm PNM}$ is the total energy of pure neutron matter. Although it is different from the usual EOS parameterization~\cite{PhysRevC.75.015801}, $E_{\rm P}$ may well be related to the symmetry energy.
\begin{figure}[t]
    \centering
    \includegraphics[width=8cm]{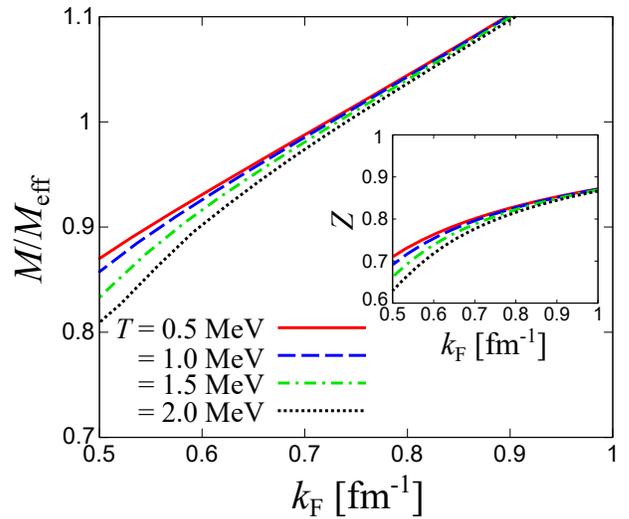}
    \caption{Inverse effective mass $M/M_{\rm eff}$ and residue $Z$ are plotted in the attractive polaron branch of an impurity proton embedded in neutron matter of Fermi momentum $k_{\rm F}$.}
    \label{fig:3}
\end{figure}

Figure~\ref{fig:3} shows the effective mass $M_{\rm eff}$ and the polaron residue $Z$ in the attractive polaron branch.
In a relatively low-density regime, $M_{\rm eff}$ is larger than $M$ as 
in the case of atomic Fermi polarons. 
This is associated with the strong neutron-proton attraction described by $\Gamma_{\sigma\sigma'}(\bm{k},\bm{k}';\bm{q},i\nu_\ell)$ in Eq.~(\ref{eq:sigma}) as well as the reduction of $Z$.
One can confirm from Fig.~\ref{fig:3}
that the results for the polaron properties are insensitive to the temperature, particularly at the highest densities considered here.
We note that $M/M_{\rm eff}\gesim 1$ is found at sufficiently high densities ($k_{\rm F}\gesim 0.8$ fm$^{-1}$) because of the finite-range properties. Indeed, this behavior can be qualitatively understood by the lowest-order shift as shown in the Supplement~\cite{Supplement}.
The reduction of $M_{\rm eff}$ in such a high-density regime is also consistent with the behavior of the Landau effective mass of Fermi-degenerate protons in asymmetric nuclear matter~\cite{li2018nucleon}.
Either way, the tendency that the lower the density, the larger the effective mass is the consequence of strong neutron-proton correlations.

\noindent{\it Diproton clustering}---
Let us now consider the possible presence of a stable diproton in neutron matter of such densities as to dissociate deuterons.
In addition to the large $M_{\rm eff}$, 
polaronic protons involve a neutron-mediated attraction associated with the spin-triplet neutron-proton interaction. 
As will be shown below, these medium effects act to stabilize diprotons, which are known to be unbound in vacuum.

We first describe the direct proton-proton interaction by
the separable spin-singlet potential
$V_{\rm s}(\bm{k},\bm{k}')=-\eta_k\eta_{k'}$ with the form factor $\eta_k=u_{\rm s}/(k^2+\Lambda_{\rm s}^2)$. Here,
$u_{\rm s}$ and $\Lambda_{\rm s}$ are determined in such a way as to reproduce the empirical proton-proton scattering length $a_{\rm s}=-17.164$~fm and effective range $r_{\rm s}=2.865$~fm that have effects of the electromagnetic interaction subtracted out~\cite{PhysRevC.51.38}.
On top of this attraction, we include the neutron-mediated interaction $V_{\rm med.}$ between two protons at rest
by replacing $\eta_{k}$ with the effective form factor $\tilde{\eta}_k=\tilde{u}_{\rm s}/(k^2+\Lambda_{\rm s}^2)$ so as to satisfy $\tilde{\eta}_{0}^2=\eta_0^2-V_{\rm med.}$.
For simplicity, $V_{\rm med.}$ has been evaluated at $T=0$ in the leading order of $V_{\rm t}$ and in the limit of zero momentum transfer as $-\frac34\gamma_{k_{\rm F}/2}^4(Mk_{\rm F}/2\pi^2)$, which is negative (attractive) for any value of $k_{\rm F}$.
We finally obtain the diproton bound-state equation from the pole of the proton-proton scattering $T$-matrix~\cite{tajima2019superfluid} as
\begin{align}
\label{eq:pp-pole}
    &1-Z^2
    \sum_{\bm{q}}\frac{\theta(k_{\rm F}-q)M_{\rm eff}\tilde{\eta}_q^2}{q^2+M_{\rm eff}E_{\rm PP}}=0,
\end{align}
where $E_{\rm PP}$ is the diproton binding energy (see Supplement~\cite{Supplement}).
Here, we have ignored $\Gamma_{\rm P}$ in Eq.~\eqref{eq:pp-pole} and assumed that the quasiparticle picture is valid for diproton relative momenta of up to $q=k_{\rm F}$~\cite{PhysRevResearch.2.023152}. Also, we have omitted contributions of virtual diprotons of $q>k_{\rm F}$ because the direct and medium-induced interactions leading to $q>k_{\rm F}$ are suppressed by the form factor $\eta_q$ and the neutron Fermi degeneracy, respectively.  

\begin{figure}[t]
    \centering
    \includegraphics[width=8cm]{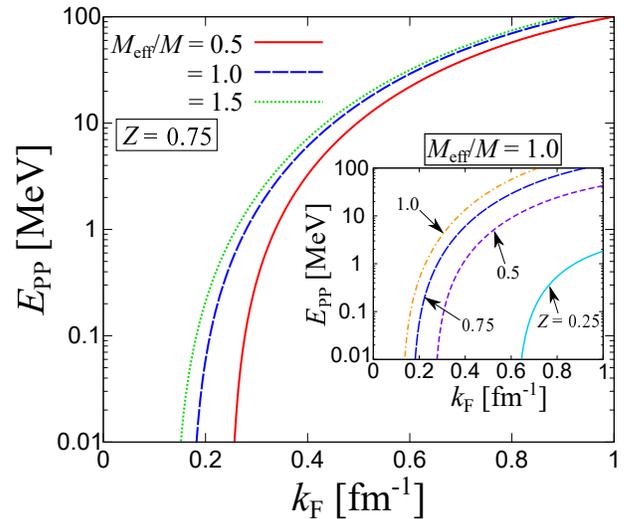}
    \caption{Diproton binding energy $E_{\rm PP}$ obtained by solving Eq.~(\ref{eq:pp-pole}) with $Z=0.75$ and $M_{\rm eff}/M=0.5$, $1.0$, and $1.5$ in neutron-rich matter at subnuclear densities.
    The inset shows $Z$-dependence of $E_{\rm PP}$ with $M_{\rm eff}/M=1.0$ fixed. 
    }
    \label{fig:4}
\end{figure}
Figure~\ref{fig:4} presents
$E_{\rm PP}$ as function of $k_{\rm F}$ for various sets of the polaron parameters $Z$ and $M_{\rm eff}$.
$E_{\rm PP}$ starts to be significant at a critical density, which is typically $k_{\rm F}\simeq 0.3$ fm$^{-1}$ but dependent on $M_{\rm eff}$ and $Z$. 
Larger $M_{\rm eff}$
leads to stronger diproton binding, 
while smaller $Z$ tends to suppress the binding substantially.
The TMA result, $Z=0.6$--$1$, shown in Fig.~\ref{fig:3} suggests that diproton binding occurs with the help of the induced interaction $V_{\rm med.}$.

The possible presence of bound diprotons in neutron matter at subnuclear densities could make a significant difference in the properties of neutron-rich nuclei in laboratories and also neutron-rich matter around the neutron star crust-core interface. This is because if we switch from a proton few-body system to a proton many-body system without changing neutron matter as a main component, one can expect a diproton Bose-Einstein-condensation (BEC) and associated proton superconductivity to occur at sufficiently low temperatures. Such a switch would not drastically change the Fermi-polaron picture of each proton~\cite{Tajima_2018,PhysRevA.98.013626,PhysRevA.105.043303}.
If the proton density is increased, moreover, a BEC--BCS crossover may occur even for protons as in the case of cold atoms near the unitary limit~\cite{zwerger2011bcs,randeria2014crossover,chen2005bcs,STRINATI20181,ohashi2020bcs}.
Indeed, at sufficiently high proton density, proton Cooper pairs would take over given that the $s$-wave direct interaction for two protons is too weak to induce binding in vacuum by itself and that the induced attraction $V_{\rm med.}$ has to be counteracted by the induced repulsion due to proton polarization.  

Recall that diprotons can occur in a neutron-rich environment at subnuclear densities. Then, one can expect the presence of diprotons in the surface region of neutron-rich nuclei both in laboratories and in neutron star crusts, as well as in uniform matter just below the crust.  However, it is not obvious whether or not such diprotons condense.  Very naively, 
the diproton BEC temperature can be estimated as $T_{\rm c}^{\rm pp}\simeq 0.218\frac{(3\pi^2\rho_{\rm p})^{2/3}}{2M_{\rm eff}}$. 
It would be interesting to consider how the existing
scenario of neutron-star cooling based on the BCS-type proton superconductivity~\cite{chamel2017superfluidity}
could be changed by the presence of diprotons.
We remark in passing that our analysis is similar to that of the nucleus-nucleus interaction in a dripped neutron gas~\cite{PhysRevC.94.055806}, which concludes that the repulsive direct interaction is canceled by the static neutron-mediated interaction.
In contrast to the latter analysis in which immiscible nuclei are assumed to be in bulk equilibrium with neutron matter, we treat protons as miscible impurities embedded locally in neutron matter and thus such a cancellation does not occur.

\noindent{\it Conclusion}---
In this work, we have clarified the polaronic properties of a proton immersed in neutron matter at subnuclear densities. 
At sufficiently high densities to satisfy $E_{\rm F}\gg E_{\rm d}$, attractive 
polarons are stable with respect to
the deuteron formation.
Due to the medium-induced attraction between two adjacent protons in the attractive polaron branch,
bound diprotons have been predicted to occur in neutron matter above a critical density corresponding to $k_{\rm F}$ of order 0.1 fm$^{-1}$.  

An alpha particle, if surviving in neutron matter above
such a critical density, may have a novel structure,
namely, a composite of a diproton and a dineutron,
which is different from the typical one of a triton and a
proton~\cite{PhysRevC.70.031001,PhysRevC.78.034305}.  The former kind of alpha clustering
is expected in the surface region of neutron-rich
nuclei in both laboratories~\cite{tanaka2021formation} and neutron star crusts. 
To obtain more certain implications of the diproton properties, many questions remain, including the effects of neutron superfluidity~\cite{PhysRevA.105.023317} and of the ignored parts of the nucleon-nucleon interactions.

\acknowledgments
H.~T. thanks H. Liang and Y. Guo for useful discussions.
This research was funded by Grants-in-Aid for Scientific
Research provided by JSPS through Nos.~18H05406, 20K14480, 22H01158, and 22K13981.

\bibliographystyle{apsrev4-2}
\bibliography{reference.bib}

%supplemental 
\clearpage

\beginsupplement

%%%%%%%%%% Merge with supplemental materials %%%%%%%%%%
%%%%%%%%%% Prefix a "S" to all equations, figures, tables and reset the counter %%%%%%%%%%
\setcounter{equation}{0}
\setcounter{figure}{0}
\setcounter{table}{0}
\setcounter{page}{1}
\makeatletter
\renewcommand{\theequation}{S\arabic{equation}}
\renewcommand{\thefigure}{S\arabic{figure}}
%\renewcommand{\bibnumfmt}[1]{[S#1]}
%\renewcommand{\citenumfont}[1]{S#1}
%%%%%%%%%% Prefix a "S" to all equations, figures, tables and reset the counter %%%%%%%%%%

\begin{widetext}
\section*{Supplemental Material: Polaronic Proton and Diproton Clustering in Neutron-Rich Matter}

\section{S1. Bethe-Salpeter equation in the spin-triplet neutron-proton channel}
\label{sup:1}
The Hamiltonian $H$ can be rewritten as
\begin{align}
H&=\sum_{\sigma=\uparrow,\downarrow}\sum_{\tau={\rm n,p}}\sum_{\bm{k}}\xi_{\bm{k},\tau} c_{\bm{k},\sigma,\tau}^\dag c_{\bm{k},\sigma,\tau}\cr
&+\sum_{\bm{k},\bm{k}',\bm{q}}\sum_{\sigma=\up,\dwn}V_{\rm t}(\bm{k},\bm{k}')
c_{\bm{  k}+\bm{q}/2,\sigma,{\rm n}}^{\dag}c_{-\bm{  k}+\bm{q}/2,\sigma,{\rm p}}^{\dag}
c_{-\bm{k}'+\bm{q}/2,\sigma,{\rm p}}c_{\bm{k}'+\bm{q}/2,\sigma,{\rm n}}\cr
&+\frac{1}{2}\sum_{\bm{k},\bm{k}',\bm{q}}V_{\rm t}(\bm{k},\bm{k}')
c_{\bm{  k}+\bm{q}/2,\up,{\rm n}}^{\dag}c_{-\bm{  k}+\bm{q}/2,\dwn,{\rm p}}^{\dag}
c_{-\bm{k}'+\bm{q}/2,\dwn,{\rm p}}c_{\bm{k}'+\bm{q}/2,\up,{\rm n}}\cr
&+\frac{1}{2}\sum_{\bm{k},\bm{k}',\bm{q}}V_{\rm t}(\bm{k},\bm{k}')
c_{\bm{  k}+\bm{q}/2,\up,{\rm p}}^{\dag}c_{-\bm{  k}+\bm{q}/2,\dwn,{\rm n}}^{\dag}
c_{-\bm{k}'+\bm{q}/2,\dwn,{\rm n}}c_{\bm{k}'+\bm{q}/2,\up,{\rm p}}\cr
&-\frac{1}{2}\sum_{\bm{k},\bm{k}',\bm{q}}V_{\rm t}(\bm{k},\bm{k}')
c_{\bm{  k}+\bm{q}/2,\up,{\rm n}}^{\dag}c_{-\bm{  k}+\bm{q}/2,\dwn,{\rm p}}^{\dag}
c_{-\bm{k}'+\bm{q}/2,\dwn,{\rm n}}c_{\bm{k}'+\bm{q}/2,\up,{\rm p}}\cr
&-\frac{1}{2}\sum_{\bm{k},\bm{k}',\bm{q}}V_{\rm t}(\bm{k},\bm{k}')
c_{\bm{  k}+\bm{q}/2,\up,{\rm p}}^{\dag}c_{-\bm{  k}+\bm{q}/2,\dwn,{\rm n}}^{\dag}
c_{-\bm{k}'+\bm{q}/2,\dwn,{\rm p}}c_{\bm{k}'+\bm{q}/2,\up,{\rm n}}.
\end{align}
Here, $\uparrow$ ($\downarrow$) stands for $\sigma=+1/2$ ($-1/2$), and
the nucleon-nucleon coupling strengths with parallel and antiparallel spins are given by $V_{\rm t}(\bm{k},\bm{k}')$
and $V_{\rm t}(\bm{k},\bm{k}')/2$, respectively.  Moreover, the isospin-exchange terms with the coupling strength $-V_{\rm t}(\bm{k},\bm{k}')/2$ 
arise.
Put more simply,
the lowest-order interaction vertex between a spin-$\sigma$ proton and a spin-$\sigma'$ neutron is
\begin{align}
   V_{\rm t}(\bm{k},\bm{k}')\delta_{\sigma,\sigma'}
    +\frac{1-\delta_{\sigma,\sigma'}}{2}V_{\rm t}(\bm{k},\bm{k}').
\label{eq:LOint}
\end{align}
To incorporate the strong-coupling effect associated with 
deuteron-like correlations, the ladder diagrams 
will be taken into account for each $S_z$ below.

\begin{figure}
    \centering
    \includegraphics[width=14cm]{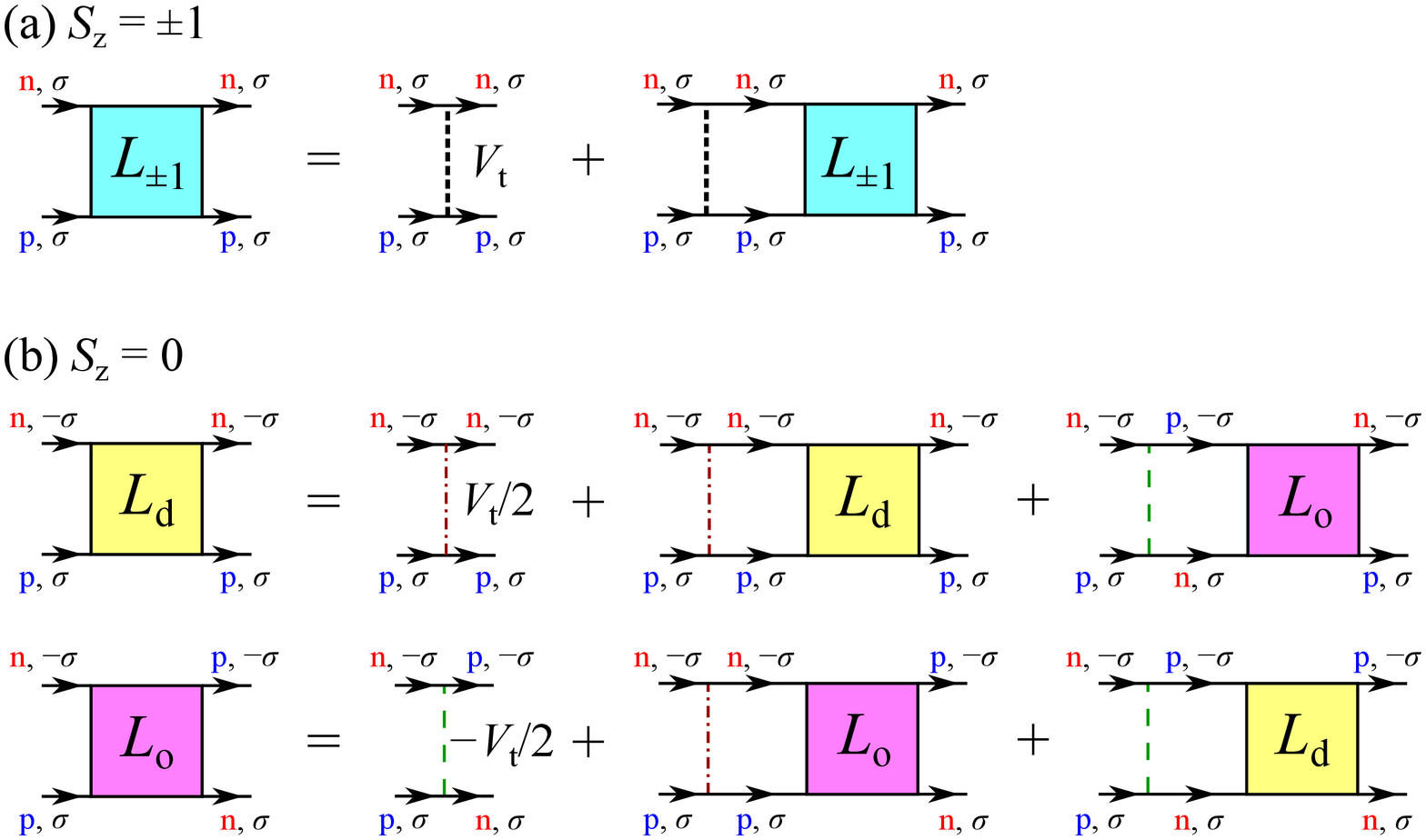}
    \caption{Diagrammatic representations of the Bethe-Salpeter equations in $S_z=\pm 1$ channel (a) and in $S_z=0$ channel (b). $L_{\pm 1}$ represents the neutron-proton vertex in the $S_z=\pm 1$ channel. $L_{\rm d}$ and $L_{\rm o}$ are the diagonal and off-diagonal neutron-proton vertices in the $S_z=0$ channel, where the off-diagonal part $L_{\rm o}$ induces the isospin exchange. The bare couplings $V_{\rm t}$ in the $S_z=\pm 1$ channel, $V_{\rm t}/2$ in the $S_z=0$ channel without isospin flip, and $- V_{\rm t}/2$ in the $S_z=0$ channel with isospin flip 
    are represented by the dotted, dotted-dashed, and dashed lines, respectively.}
    \label{fig:s1}
\end{figure}
For the $S_z=\pm 1$ channel where a neutron and a proton with parallel spins interact with each other, 
the Bethe-Salpeter equation of the neutron-proton vertex $L_{\pm 1}(\bm{k},\bm{k}';\bm{q},i\nu_{\ell})$ can be diagrammatically shown in Fig.~\ref{fig:s2}(a).  Its explicit form reads
\begin{align}
    &L_{\pm 1}(\bm{k},\bm{k}';\bm{q},i\nu_{\ell})
    =V_{\rm t}(\bm{k},\bm{k}')\cr
    &\quad\quad
    -T\sum_{\bm{k}'',i\omega_n}V_{\rm t}(\bm{k},\bm{k}'')
    G_{\rm n}^0(\bm{k}''+\bm{q}/2,i\omega_n+i\nu_\ell)
    G_{\rm p}^0(-\bm{k}''+\bm{q}/2,-i\omega_n)L_{\pm 1}(\bm{k},\bm{k}';\bm{q},i\nu_{\ell}),
\end{align}
where $G_{\rm n,p}^0(\bm{k},i\omega_n)=(i\omega_n-\xi_{\bm{k},{\rm n,p}})^{-1}$ is the thermal Green's function of a bare nucleon.
Assuming the separable interaction $V_{\rm t}(\bm{k},\bm{k}')=-\gamma_k\gamma_{k'}$,
we introduce $L_{\pm 1}(\bm{k},\bm{k}';\bm{q},i\nu_{\ell})=L_{1}(\bm{q},i\nu_\ell)\gamma_k\gamma_{k'}$.
Here, $L_1(\bm{q},i\nu_\ell)$ can be obtained as
\begin{align}
L_1(\bm{q},i\nu_\ell)=-1+\Pi(\bm{q},i\nu_\ell)L_1(\bm{q},i\nu_\ell)=-\frac{1}{1-\Pi(\bm{q},i\nu_\ell)},
\end{align}
where 
\begin{align}
\Pi(\bm{q},i\nu_\ell)&=T\sum_{\bm{k},i\omega_n}\gamma_k^2G_{\rm n}^0(\bm{k}+\bm{q}/2,i\omega_n+i\nu_\ell)G_{\rm p}^0(-\bm{k}+\bm{q}/2,-i\omega_n)\cr
%    &=-T\sum_{\bm{k},i\omega_n}\gamma_k^2
%    \frac{1}{i\omega_n+i\nu_\ell-\xi_{\bm{k}+\bm{q}/2,{\rm n}}}\frac{1}{i\omega_n+\xi_{-\bm{k}+\bm{q}/2,{\rm p}}}\cr
%    &=\sum_{\bm{k}}\oint\frac{f(z)dz}{2\pi i}\gamma_k^2
%    \frac{1}{z+i\nu_\ell-\xi_{\bm{k}+\bm{q}/2,{\rm n}}}\frac{1}{z+\xi_{-\bm{k}+\bm{q}/2,{\rm p}}}\cr
%    &=-\sum_{\bm{k}}\gamma_k^2
%    \left[\frac{f(-\xi_{-\bm{k}+\bm{q}/2,{\rm p}})}{i\nu_\ell-\xi_{\bm{k}+\bm{q}/2,{\rm n}}-\xi_{-\bm{k}+\bm{q}/2,{\rm p}}}+
%  \frac{f(\xi_{\bm{k}+\bm{q}/2,{\rm n}})}{-i\nu_\ell+\xi_{\bm{k}+\bm{q}/2,{\rm n}}+\xi_{-\bm{k}+\bm{q}/2,{\rm p}}}
%    \right]
%    \cr
    &=-\sum_{\bm{k}}\gamma_k^2\frac{1-f(\xi_{-\bm{k}+\bm{q}/2,{\rm p}})-f(\xi_{\bm{k}+\bm{q}/2,{\rm n}})}{i\nu_\ell-\xi_{\bm{k}+\bm{q}/2,{\rm n}}-\xi_{-\bm{k}+\bm{q}/2,{\rm p}}}
\end{align}
is the neutron-proton propagator.
Because the dilute limit is considered for protons, the proton distribution function is taken to be zero [i.e., $f(\xi_{\bm{k},{\rm p}})\rightarrow 0$ with $\mu_{\rm p}\rightarrow -\infty$].

For the $S_z=0$ channel, which is open for two nucleons with antiparallel spins, the isospin exchange can occur due to the spin-triplet scattering. %for two nucleons with opposite spins.
Indeed, as shown diagrammatically in Fig.~\ref{fig:s2}(b), we can express the Bethe-Salpeter equations for the diagonal part without isospin flip accompanied by the off-diagonal part with isospin flip and vice versa as
\begin{align}
    L_{\rm d}(\bm{k},\bm{k}';\bm{q},i\nu_\ell)=\frac{V_{\rm t}(\bm{k},\bm{k}')}{2}&-T\sum_{\bm{k}'',i\omega_{n}}\frac{V_{\rm t}(\bm{k},\bm{k}'')}{2}
    G_{\rm n}^0(\bm{k}''+\bm{q}/2,i\omega_n+i\nu_\ell)
    G_{\rm p}^0(-\bm{k}''+\bm{q}/2,-i\omega_n)
    L_{\rm d}(\bm{k}'',\bm{k}';\bm{q},i\nu_\ell)\cr
    &+T\sum_{\bm{k}'',i\omega_{n}}\frac{V_{\rm t}(\bm{k},\bm{k}'')}{2}G_{\rm n}^0(\bm{k}''+\bm{q}/2,i\omega_n+i\nu_\ell)
    G_{\rm p}^0(-\bm{k}''+\bm{q}/2,-i\omega_n)
    L_{\rm o}(\bm{k}'',\bm{k}';\bm{q},i\nu_\ell)
\end{align}
and 
%the off-diagonal part
\begin{align}
    L_{\rm o}(\bm{k},\bm{k}';\bm{q},i\nu_\ell)=-\frac{V_{\rm t}(\bm{k},\bm{k}')}{2}&+T\sum_{\bm{k}'',i\omega_{n}}\frac{V_{\rm t}(\bm{k},\bm{k}'')}{2}G_{\rm n}^0(\bm{k}''+\bm{q}/2,i\omega_n+i\nu_\ell)
    G_{\rm p}^0(-\bm{k}''+\bm{q}/2,-i\omega_n)
    L_{\rm d}(\bm{k}'',\bm{k}';\bm{q},i\nu_\ell)\cr
    &-T\sum_{\bm{k}'',i\omega_{n}}\frac{V_{\rm t}(\bm{k},\bm{k}'')}{2}G_{\rm n}^0(\bm{k}''+\bm{q}/2,i\omega_n+i\nu_\ell)
    G_{\rm p}^0(-\bm{k}''+\bm{q}/2,-i\omega_n)
    L_{\rm o}(\bm{k}'',\bm{k}';\bm{q},i\nu_\ell),
\end{align}
respectively.  Here,
one can easily find $L_{\rm o}(\bm{k},\bm{k}';\bm{q},i\nu_\ell)=-L_{\rm d}(\bm{k},\bm{k}';\bm{q},i\nu_\ell)$.
%In this regard,
For the separable interaction, in a manner similar to the $S_z=\pm 1$ channel,
we introduce $L_{\rm d}(\bm{k},\bm{k}';\bm{q},i\nu_\ell)=-L_{\rm o}(\bm{k},\bm{k}';\bm{q},i\nu_\ell)\equiv L_0(\bm{q},i\nu_\ell)\gamma_k\gamma_{k'}$.
We can then obtain
\begin{align}
    L_0(\bm{q},i\nu_\ell)&=-\frac{1}{2}
    +\Pi(\bm{q},i\nu_\ell)L_0(\bm{q},i\nu_\ell)=-\frac{1}{2\left[1-\Pi(\bm{q},i\nu_\ell)\right]}.
\end{align}

Finally, we introduce a unified form of the in-medium neutron-proton $T$-matrix according to 
\begin{align}
    \Gamma_{\sigma\sigma'}(\bm{k},\bm{k}';\bm{q},i\nu_\ell)&=
    \left[L_{1}(\bm{q},i\nu_\ell)\delta_{\sigma,\sigma'}
    +L_{0}(\bm{q},i\nu_\ell)\left(1-\delta_{\sigma,\sigma'}\right)\right]\gamma_k\gamma_{k'}\cr
    &=-\frac{\gamma_{k}\gamma_{k}'}{1-\Pi(\bm{q},i\nu_\ell)}\left(\delta_{\sigma,\sigma'}+\frac{1-\delta_{\sigma,\sigma'}}{2}\right),
\end{align}
which is used in the main text.

\section{S2. Self-energy for a polaronic proton within the many-body $T$-matrix approach}
We consider the thermal proton Green's function
\begin{align}
    G_{{\rm p}\sigma}^T(\bm{k},i\omega_n)
    =\frac{1}{i\omega_n-\xi_{\bm{k},{\rm p}}-    \Sigma_{{\rm p}\sigma}(\bm{k},i\omega_n)}
\end{align}
where $\omega_n=(2n+1)\pi T$ is the fermion Matsubara frequency ($n\in \mathbb{Z}$).
The proton self-energy within the finite-temperature TMA 
is given by
\begin{align}
    \Sigma_{{\rm p}\sigma}^{T}(\bm{k},i\omega_n)
    &=T\sum_{\bm{q}}\sum_{\sigma'}\sum_{i\nu_\ell}
    \Gamma_{\sigma\sigma'}(\bm{q}/2-\bm{k},\bm{q}/2-\bm{k};\bm{q},i\nu_\ell) G_{{\rm n}\sigma'}^T(\bm{q}-\bm{k},i\nu_\ell-i\omega_n),
\end{align}
where $\nu_\ell=2\ell \pi T$ is the boson Matsubara frequency ($\ell \in \mathbb{Z}$).
In the dilute limit of protons, the backaction on the medium (i.e., uniform neutron matter) can be neglected, so that the thermal neutron Green's function reads $G_{{\rm n}\sigma}^T(\bm{q}-\bm{k},i\nu_\ell-i\omega_n)\simeq G_{\rm n}^0(\bm{q}-\bm{k},i\nu_\ell-i\omega_n)\equiv \left(i\nu_\ell-i\omega_n-\xi_{\bm{q}-\bm{k},{\rm n}}\right)^{-1}$.
The summation over the boson Matsubara frequency can be replaced by the contour integral enclosing the pole of the Bose distribution function $b(z)=\left[e^{z/T}-1\right]^{-1}$ as
\begin{align}
        \Sigma_{{\rm p}\sigma}^{T}(\bm{k},i\omega_n)
    &=-\sum_{\bm{q}}\sum_{\sigma'}\oint\frac{b(z)dz}{2\pi i}
    \frac{\Gamma_{\sigma\sigma'}(\bm{q}/2-\bm{k},\bm{q}/2-\bm{k};\bm{q},z)}{z-i\omega_n-\xi_{\bm{q}-\bm{k},{\rm n}}}.
\end{align}
The contour integral in $\Sigma_{{\rm p}\sigma}^T(\bm{k},i\omega_n)$ involves two poles $z_1$ and $z_2$ that satisfy $z_1=i\omega_n+\xi_{\bm{q}-\bm{k},{\rm n}}$ and $\Gamma_{\sigma\sigma'}^{-1}(\bm{q}/2-\bm{k},\bm{q}/2-\bm{k};\bm{q},z_2)=0$,  respectively.
The latter one gives the contribution proportional to the bosonic distribution $b(\varepsilon)$ where $\varepsilon$ is the deuteron-like excitation energy.
Because of the proton dilute limit considered here, the bosonic distribution can be ignored.
In this way, the fermionic pole $z_1$
%$\left(z-i\omega_n-\xi_{\bm{q}-\bm{k},{\rm n}}\right)^{-1}$ 
leads to
\begin{align}
        \Sigma_{{\rm p}\sigma}^T(\bm{k},i\omega_n)
        =\sum_{\bm{q}}\sum_{\sigma'}\Gamma_{\sigma\sigma'}(\bm{q}/2-\bm{k},\bm{q}/2-\bm{k};\bm{q},i\omega_n+\xi_{\bm{q}-\bm{k},{\rm n}})f(\xi_{\bm{q}-\bm{k},{\rm n}}).
\label{eq:selfTMA}
\end{align}
Finally, the retarded self-energy $\Sigma_{{\rm p}\sigma}(\bm{k},\omega)$ can be obtained via the analytic continuation $i\omega_n\rightarrow\omega_+-\mu_{\rm p}$ [i.e., $\Sigma_{{\rm p}\sigma}(\bm{k},\omega)\equiv\Sigma_{{\rm p}\sigma}^T(\bm{k},i\omega_n\rightarrow\omega_+-\mu_{\rm p})$].

To extract the polaronic properties, we expand the self-energy around $\bm{k}=0$ and $\omega=E_{\rm P}$ (where $E_{\rm P}$ is the polaron energy) as
\begin{align}
    \Sigma_{{\rm p}\sigma}(\bm{k},\omega)\simeq
    \Sigma_{{\rm p}\sigma}(\bm{0},E_{\rm P})+\left.\frac{\partial \Sigma_{{\rm p}\sigma}(\bm{k},\omega)}{\partial \omega}\right|_{\omega=E_{\rm P}}(\omega-E_{\rm P})+\frac{1}{2}\left.\frac{\partial^2 \Sigma_{{\rm p}\sigma}(\bm{k},E_{\rm P})}{\partial k^2}\right|_{\bm{k}=0}k^2,
\end{align}
which leads to
\begin{align}
    G_{{\rm p}\sigma}(\bm{k},\omega)&\simeq\left[\omega+i\delta-\frac{k^2}{2M}-\Sigma_{{\rm p}\sigma}(\bm{0},E_{\rm P})
    -\left.\frac{\partial \Sigma_{{\rm p}\sigma}(\bm{k},\omega)}{\partial \omega}\right|_{\omega=E_{\rm P}}(\omega-E_{\rm P})
    -\frac{1}{2}\left.\frac{\partial^2 \Sigma_{{\rm p}\sigma}(\bm{k},E_{\rm P})}{\partial k^2}\right|_{\bm{k}=0}k^2\right]^{-1}\cr
    &\equiv\frac{Z}{\omega-\frac{k^2}{2M_{\rm eff}}-E_{\rm P}+i\Gamma_{\rm P}/2}.
\end{align}
We therefrom obtain the polaron energy
\begin{align}
    E_{\rm P}={\rm Re}\Sigma_{{\rm p}\sigma}(\bm{0},E_{\rm P}),
\end{align}
the polaron residue
\begin{align}
    Z=\left[1-{\rm Re}\left.\frac{\partial \Sigma_{{\rm p}\sigma}(\bm{0},\omega)}{\partial \omega}\right|_{\omega=E_{\rm P}}\right]^{-1},
\end{align}
the inverse effective mass
\begin{align}
    \frac{M}{M_{\rm eff}}=Z\left[1+M{\rm Re}\left.\frac{\partial^2 \Sigma_{{\rm p}\sigma}(\bm{k},E_{\rm P})}{\partial k^2}\right|_{\bm{k}=\bm{0}}\right].
\end{align}
and the decay rate
\begin{align}
    \Gamma_{\rm P}=-2Z{\rm Im}\Sigma_{{\rm p}\sigma}(\bm{0},E_{\rm P}).
\end{align}
%are introduced.

The spectral weight of a zero-momentum polaronic proton is defined as
\begin{align}
    A_{\rm p}(\omega)=-\frac{1}{\pi}{\rm Im}G_{{\rm p}\sigma}(\bm{k}=\bm{0},\omega).
\end{align}
We note that $A_{\rm p}(\omega)$ does not have a $\sigma$ dependence because we consider a spin-unpolarized neutron system.
\begin{figure}
    \centering
    \includegraphics[width=8cm]{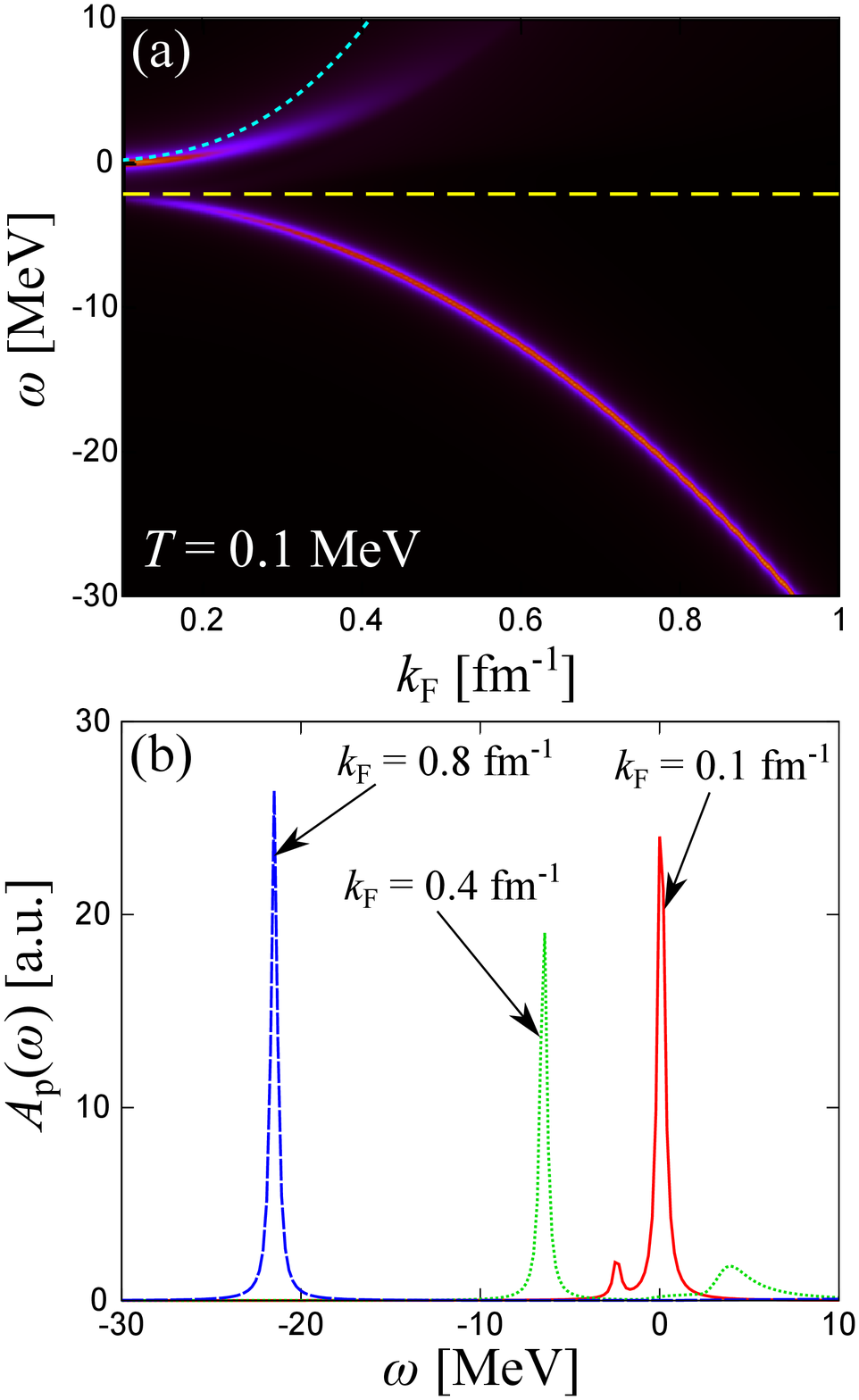}
    \caption{(a) Spectral weight $A_{\rm p}(\omega)$ of a polaronic proton at $T=0.1$ MeV. 
     The color brightness indicates the intensity of $A_{\rm p}(\omega)$ in an arbitrary unit. The dotted line shows the Hartree shift, while the dashed line is the deuteron energy $-E_{\rm d}=-2.2$ MeV.
    The panel (b) shows $A_{\rm p}(\omega)$ at $k_{\rm F}=0.1$ fm$^{-1}$, $0.4$ fm$^{-1}$, and $0.8$ fm$^{-1}$.  }
    \label{fig:s2}
\end{figure}
In addition to Fig.~2 that exhibits $A_{{\rm p}}(\omega)$ at $T=2.0$ MeV,
we show $A_{{\rm p}}(\omega)$ at $T=0.1$ MeV in Fig.~\ref{fig:s2}.
Such a low-temperature result is more relevant to neutron-rich nuclei in laboratories and neutron stars, although
neutron-neutron pairing ignored here 
would make a quantitative difference in the polaron properties~\cite{PhysRevA.105.023317}.
One can find that the spectral properties are qualitatively similar to the result at $T=2.0$ MeV.
In contrast to the case of $T=2.0$ MeV, however, the attractive polaron branch looks sharp even at low neutron density, and also the repulsive polaron branch extends down to a very low neutron density. 

\begin{figure}
    \centering
    \includegraphics[width=8cm]{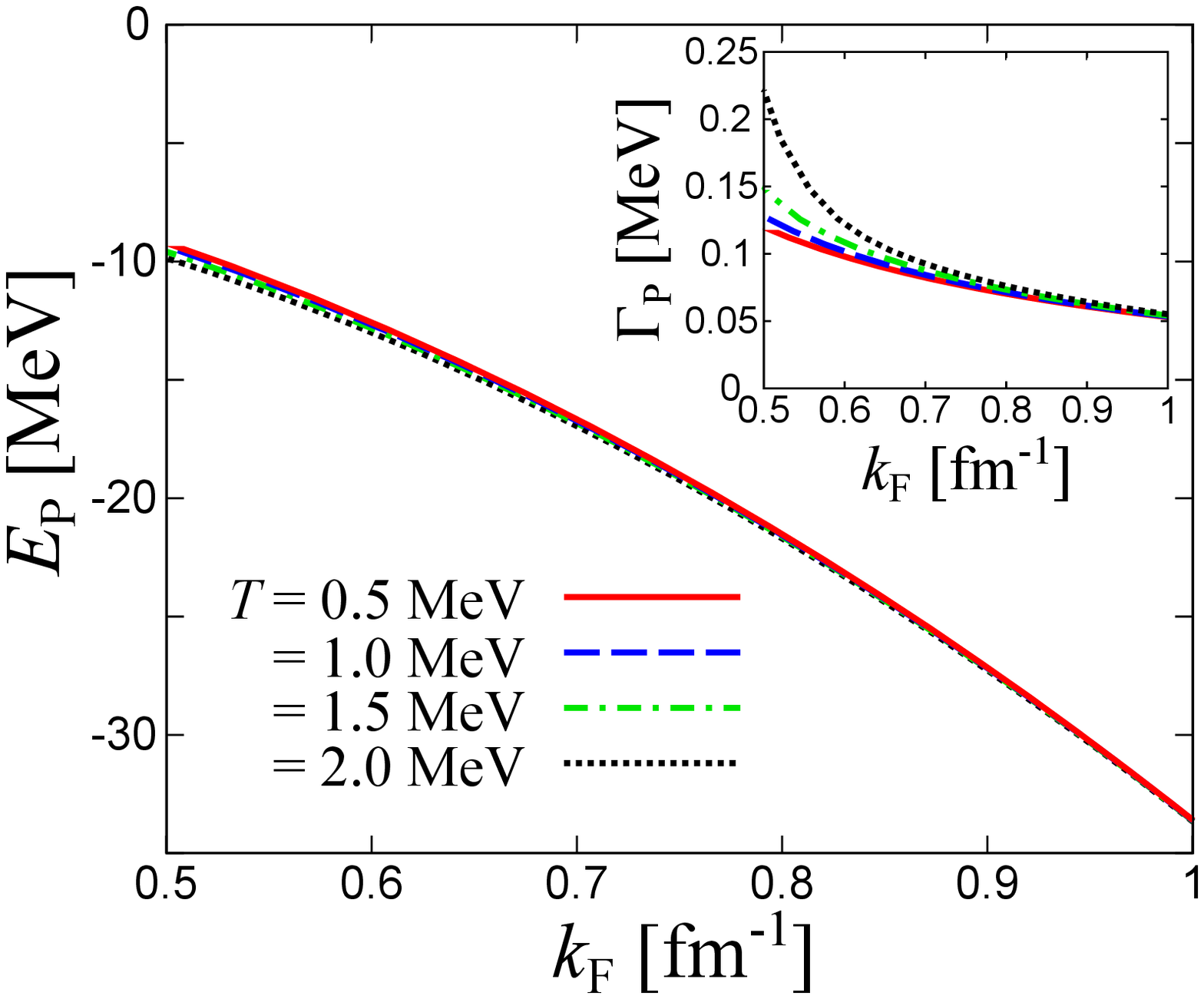}
    \caption{Polaron energy $E_{\rm P}$ in the attractive branch extracted from the proton Green's function $G_{{\rm p}\sigma}(\bm{k},\omega)$. The inset shows the decay rate $\Gamma_{\rm P}$. }
    \label{fig:s3}
\end{figure}
Figure~\ref{fig:s3} shows the polaron energy $E_{\rm P}$ in the attractive branch at different temperatures. One can see that $-E_{\rm P}$ is sufficiently larger than the decay rate $\Gamma_{\rm P}$, indicating that the attractive polaron is stabilized against the deuteron formation. Moreover, the results are relatively insensitive to the change of temperature in the Fermi-degenerate regime.

\section{S3. Lowest-order approximation to the proton self-energy}
\label{sup:2}
The lowest-order shift, which can be obtained by replacing $\Gamma_{\sigma\sigma'}$ by Eq.\ (\ref{eq:LOint}) in Eq.\ (\ref{eq:selfTMA}), reads
\begin{align}
    \Sigma_{\rm p\sigma}^{\rm LO}(\bm{k})=-\sum_{\bm{q}}\sum_{\sigma}
    \left(\frac{1+\delta_{\sigma,\sigma'}}{2}\right)
    V_{\rm t}\left(\frac{\bm{k}-\bm{q}}{2},\frac{\bm{k}-\bm{q}}{2}\right)f(\xi_{\bm{q},{\rm n}}).
\end{align}
At zero temperature, we obtain
\begin{align}
    \Sigma_{{\rm p}\sigma}^{\rm LO}(\bm{k})&=-\frac{3}{2}\sum_{\bm{q}}\gamma_{\frac{|\bm{k}-\bm{q}|}{2}}^2f(\xi_{\bm{q},{\rm n}})\cr
    %&=-\frac{3}{2}\frac{1}{(2\pi)^3}
    %\int_0^{2\pi}d\phi
    %\int_{-1}^{1}ds
    %\int_0^{k_{\rm F}}q^2 dq
    %\gamma_{\frac{\sqrt{k^2+q^2+2kqs}}{2}}^2 \cr
    %    &=-\frac{3u^2}{8\pi^2}
    %\int_{-1}^{1}ds
    %\int_0^{k_{\rm F}}q^2 dq
    %\frac{1}{\left(\Lambda^2+\frac{k^2+q^2+2kqs}{4}\right)^2}\cr
    %&=-\frac{3u^2}{8\pi^2}
    %\int_0^{k_{\rm F}}q^2 dq
    %\left[-\frac{1}{\frac{kq}{2}\left(\Lambda^2+\frac{k^2+q^2+2kqs}{4}\right)}\right]_{-1}^{1}\cr
    %    &=-\frac{3u^2}{4\pi^2 k}
    %\int_0^{k_{\rm F}}q dq
    %\left[\frac{1}{\Lambda^2+\frac{k^2+q^2-2kq}{4}}-\frac{1}{\Lambda^2+\frac{k^2+q^2+2kq}{4}}\right]\cr
    %&=-\frac{3u^2}{\pi^2 k}
    %\int_0^{k_{\rm F}}q dq
    %\left[\frac{1}{q^2-2kq+4\Lambda^2+k^2}-\frac{1}{q^2+2kq+4\Lambda^2+k^2}\right]\cr
    %&=-\frac{3u^2}{\pi^2 k}
    %\left[\frac{1}{2}\ln(q^2-2kq+4\Lambda^2+k^2)
    %-\frac{1}{2}\ln(q^2+2kq+4\Lambda^2+k^2)\right.\cr
    %&\left.
    %-\frac{-2k\tan^{-1}\left(\frac{-2k+2q}{\sqrt{16\Lambda^2+4k^2-4k^2}}\right)}{\sqrt{16\Lambda^2+4k^2-4k^2}}
    %+\frac{2k\tan^{-1}\left(\frac{2k+2q}{\sqrt{16\Lambda^2+4k^2-4k^2}}\right)}{\sqrt{16\Lambda^2+4k^2-4k^2}}
    %\right]_{0}^{k_{\rm F}}\cr
    &=-\frac{3u_{\rm t}^2}{2\pi^2}
    \left[\frac{1}{k}\ln\left(\frac{k_{\rm F}^2-2kk_{\rm F}+4\Lambda_{\rm t}^2+k^2}{k_{\rm F}^2+2kk_{\rm F}+4\Lambda_{\rm t}^2+k^2}\right)
    +\frac{\tan^{-1}\left(\frac{k_{\rm F}-k}{2\Lambda_{\rm t}}\right)}{\Lambda_{\rm t}}
    +\frac{\tan^{-1}\left(\frac{k_{\rm F}+k}{2\Lambda_{\rm t}}\right)}{\Lambda_{\rm t}}
    \right].
\end{align}
%where
%\begin{align}
%    \int ds\frac{1}{\left(\Lambda^2+\frac{k^2+q^2+2kqs}{4}\right)^2}
%    =-\frac{1}{\frac{2kq}{4}\left(\Lambda^2+\frac{k^2+q^2+2kqs}{4}\right)}+{\rm const.}
%\end{align}
%\begin{align}
%    \int\frac{qdq}{q^2\pm 2kq+4\Lambda^2+k^2}=
%    \frac{1}{2}\ln(q^2\pm 2kq+4\Lambda^2+k^2)
%    \mp\frac{k}{2\Lambda}\tan^{-1}\left(\frac{q\pm k}{2\Lambda}\right)+{\rm const.}
%\end{align}
Furthermore, we expand this result with respect to $k$ as
\begin{align}
    \Sigma_{{\rm p}\sigma}^{\rm LO}(\bm{k})
    %&=-\frac{3u^2}{2\pi^2 k_{\rm F}}
    %\left[\frac{2k_{\rm F}}{\Lambda}\tan^{-1}\left(\frac{k_{\rm F}}{2\Lambda}\right)-\frac{4}{1+\left(\frac{2\Lambda}{k_{\rm F}}\right)^2}\right]\cr
    %&-\frac{3u^2}{2\pi^2 k_{\rm F}}\left(\frac{k}{k_{\rm F}}\right)^2
    %\left[\frac{4\left(\frac{k_{\rm F}}{2\Lambda}\right)^4-\frac{4}{3}\left(\frac{k_{\rm F}}{2\Lambda}\right)^{6}}{\left[1+\left(\frac{k_{\rm F}}{2\Lambda}\right)^2
    %\right]^3}
    %-\frac{4\left(\frac{k_{\rm F}}{2\Lambda}\right)^4\left[1+\left(\frac{k_{\rm F}}{2\Lambda}\right)^2\right]}{\left[1+\left(\frac{k_{\rm F}}{2\Lambda}\right)^2\right]^3}
    %\right]+O(k^4)\cr
    %&=-\frac{3u^2}{2\pi^2 k_{\rm F}}
    %\left[4\frac{k_{\rm F}}{2\Lambda}\tan^{-1}\left(\frac{k_{\rm F}}{2\Lambda}\right)-\frac{4\left(\frac{k_{\rm F}}{2\Lambda}\right)^2}{1+\left(\frac{k_{\rm F}}{2\Lambda}\right)^2}\right]\cr
    %&-\frac{3u^2}{2\pi^2 k_{\rm F}}\left(\frac{k}{k_{\rm F}}\right)^2
    %\left[\frac{4\left(\frac{k_{\rm F}}{2\Lambda}\right)^4-\frac{4}{3}\left(\frac{k_{\rm F}}{2\Lambda}\right)^{6}-4\left(\frac{k_{\rm F}}{2\Lambda}\right)^4-4\left(\frac{k_{\rm F}}{2\Lambda}\right)^6}{\left[1+\left(\frac{k_{\rm F}}{2\Lambda}\right)^2
    %\right]^3}
    %\right]+O(k^4)\cr
    &=-\frac{6u_{\rm t}^2}{\pi^2 k_{\rm F}}
    \left[\frac{k_{\rm F}}{2\Lambda_{\rm t}}\tan^{-1}\left(\frac{k_{\rm F}}{2\Lambda_{\rm t}}\right)-\frac{\left(\frac{k_{\rm F}}{2\Lambda_{\rm t}}\right)^2}{1+\left(\frac{k_{\rm F}}{2\Lambda_{\rm t}}\right)^2}\right]
    +\frac{8u_{\rm t}^2}{\pi^2 k_{\rm F}}\left(\frac{k}{k_{\rm F}}\right)^2
    \frac{\left(\frac{k_{\rm F}}{2\Lambda_{\rm t}}\right)^{6}}{\left[1+\left(\frac{k_{\rm F}}{2\Lambda_{\rm t}}\right)^2
    \right]^3}
    +O(k^4).
\end{align}
In this way, the effective mass within the lowest-order approximation reads
%\begin{align}
%    \left.\frac{\partial \Sigma_{{\rm p}\sigma}^{\rm H}(\bm{k})}{\partial k^2}
%    \right|_{\bm{k}=\bm{0}}
%    &=\frac{16u^2}{\pi^2 k_{\rm F}^3}
%    \frac{\left(\frac{k_{\rm F}}{2\Lambda}\right)^{6}}{\left[1+\left(\frac{k_{\rm F}}{2\Lambda}\right)^2
%    \right]^3},
%\end{align}
\begin{align}
    \frac{M}{M_{\rm eff}^{\rm LO}}
    =1+
    \frac{16Mu_{\rm t}^2}{\pi^2 k_{\rm F}^3}
    \frac{\left(\frac{k_{\rm F}}{2\Lambda_{\rm t}}\right)^{6}}{\left[1+\left(\frac{k_{\rm F}}{2\Lambda_{\rm t}}\right)^2
    \right]^3}.
\end{align}

\begin{figure}
    \centering
    \includegraphics[width=8cm]{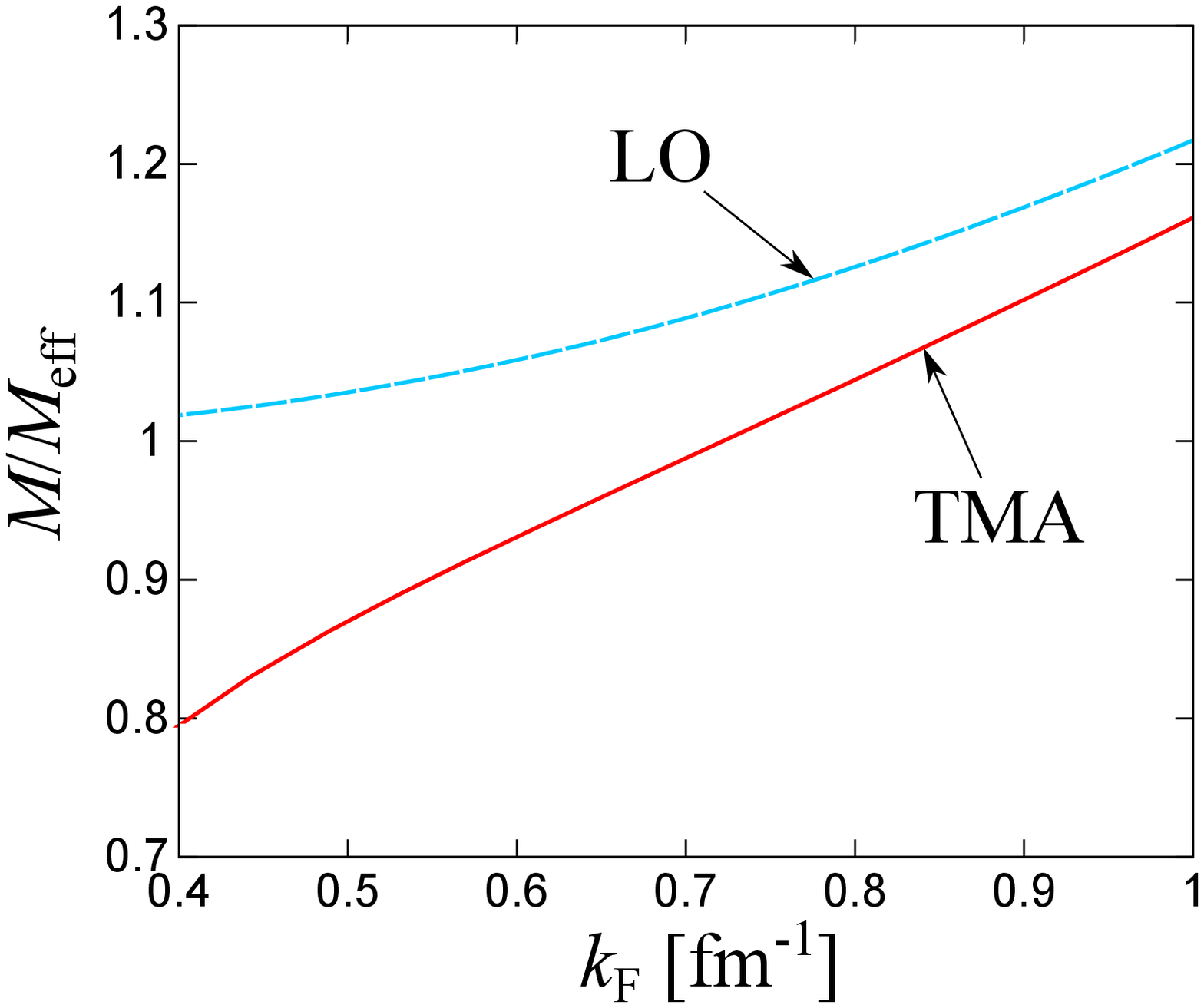}
    \caption{Inverse effective mass $M/M_{\rm eff}^{\rm LO}$ within the lowest-order approximation. For comparison, the TMA result at $T=0.5$~MeV is plotted.}
    \label{fig:s4}
\end{figure}
Figure~\ref{fig:s4} shows a comparison of the inverse effective mass between the lowest-order approximation and TMA, where we set $T=0.5$~MeV in the latter calculation.
%While 
Generally, the TMA result exhibits a heavier effective mass; this tendency is evident, especially at low neutron density.
%the lowest-order approximation shows the lighter effective masses in the entire density region.
In this regard, virtual molecular-state formation included only in the TMA plays an important role in increasing the effective mass.
At relatively high densities where the short-range part of the interaction becomes important, however,
the effective mass decreases due to the finite-range correction, a tendency that can be found in both approximations.

\section{S4. bound-state equation for two polaronic protons}
To clarify the fate of two adjacent polaronic protons qualitatively,
we consider the direct $^1S_0$ and indirect (neutron-mediated) proton-proton interaction phenomenologically.
As diagrammatically represented in Fig.~\ref{fig:s5},
the low-energy neutron-mediated interaction between two protons in the spin-singlet state can be obtained
%approximately given by
up to leading order in $V_{\rm t}$ and in the limit of zero momentum transfer as
\begin{align}
    V_{\rm med.}=%\left(\frac{4\pi a_{\rm t}}{M}\right)^2
    T\sum_{\bm{k},i\omega_n}
    %\gamma_{k/2}^4
    [V_{\rm t}(\bm{k}/2,\bm{k}/2)]^2
    \left[
    \frac{1}{2}
    \left\{G_{{\rm n}\uparrow}^0(\bm{k},i\omega_n)\right\}^2
    +\frac{1}{2}\left\{G_{{\rm n}\downarrow}^0(\bm{k},i\omega_n)\right\}^2
    -\frac{1}{4}
    G_{{\rm n}\uparrow}^0(\bm{k},i\omega_n)
    G_{{\rm n}\downarrow}^0(\bm{k},i\omega_n)
    \right],
\end{align}
where $G_{{\rm n}\sigma}^0(\bm{k},i\omega_n)=(i\omega_n-\xi_{\bm{k},{\rm n}})^{-1}$ is the thermal Green's function of a neutron with spin $\sigma= \uparrow (+1/2), \ \downarrow (-1/2)$.
At $T=0$, we obtain
\begin{align}
    V_{\rm med.}%=-\left(\frac{4\pi a_{\rm t}}{M}\right)^2\frac{5Mk_{\rm F}}{8\pi^2}.
    =-\frac{3}{4}\gamma_{k_{\rm F}/2}^4\frac{Mk_{\rm F}}{2\pi^2}=-\left(\frac{\frac{4\pi a_{\rm t}}{M}}{1-\frac{a_{\rm t}\Lambda_{\rm t}}{2}}\right)^2
    \frac{1}{\left(1+\frac{k_{\rm F}^2}{4\Lambda_{\rm t}^2}\right)^4}
    \frac{3Mk_{\rm F}}{8\pi^2}.
\end{align}
Here we introduce an effective interaction $V_{\rm eff}(\bm{k},\bm{k}')=-\tilde{\eta}_{k}\tilde{\eta}_{k'}$ with $\tilde{\eta}_k=\tilde{u}_{\rm s}/(k^2+\Lambda_{\rm s}^2)$
such that $V_{\rm eff}(\bm{k},\bm{k}')$ reproduces $V_{\rm s}(\bm{k},\bm{k}')+V_{\rm med.}$ 
%is reproduced 
in the low-momentum limit ($\bm{k}=\bm{k}'\rightarrow 0$) or, equivalently,
%Namely,
\begin{align}
    \frac{\tilde{u}_{\rm s}^2}{\Lambda_{\rm s}^4}=\frac{u_{\rm s}^2}{\Lambda_{\rm s}^4}+%\left(\frac{4\pi a_{\rm t}}{M}\right)^2
    %\frac{u_{\rm t}^4}{\Lambda_{\rm t}^8}
    %\left(\frac{\frac{4\pi a_{\rm t}}{M}}{1-\frac{a_{\rm t}\Lambda_{\rm t}}{2}}\right)^2
    \frac{6a_{\rm t}^2k_{\rm F}/M}{\left(1-\frac{a_{\rm t}\Lambda_{\rm t}}{2}\right)^2\left(1+\frac{k_{\rm F}^2}{4\Lambda_{\rm t}^2}\right)^4}.
\end{align}
\begin{figure}[t]
    \centering
    \includegraphics[width=14cm]{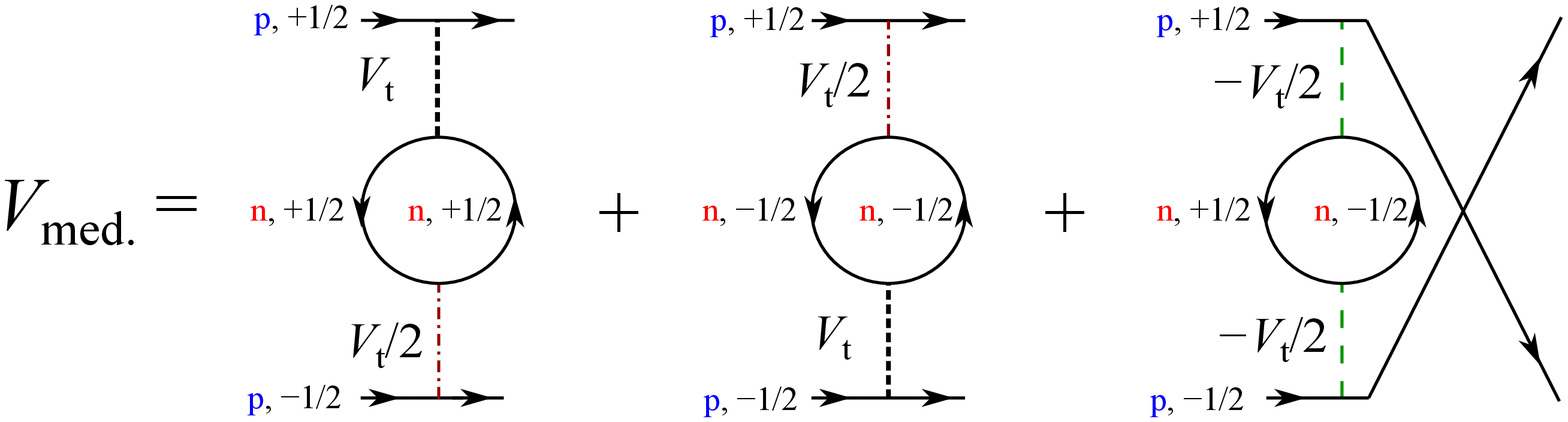}
    \caption{Feynman diagram for the neutron-mediated interaction $V_{\rm med.}$ in the spin-singlet proton-proton channel. The third diagram involves an additional minus sign compared to the others due to the spin exchange.}
    \label{fig:s5}
\end{figure}

Assuming that the well-defined quasiparticle is found only below $k=k_{\rm F}$ with sufficiently small decay rate $\Gamma_{\rm P}$,
we start from the thermal polaron propagator approximately given by
\begin{align}
    G_{\rm p\sigma}^T(\bm{k},i\omega_n)\simeq \frac{Z}{i\omega_n-\frac{k^2}{2M_{\rm eff}}-E_{\rm P}+\mu_{\rm p}}\theta(\zeta k_{\rm F}-k).
\end{align}
The parameter $\zeta$ characterizes the momentum cutoff of order $k_{\rm F}$ below which the polaron picture is valid.
%In this regard, 
In the main text, $\zeta=1$ is taken, while we examine the $\zeta$ dependence in this Supplement.
The Lippmann-Schwinger equation for the proton-proton $T$-matrix $T_{\rm pp}(\bm{k},\bm{k}';i\nu_\ell)$ at zero center-of-mass momentum (where $\nu_\ell=2\pi\ell T$ is the boson Matsubara frequency) reads
\begin{align}
    T_{\rm pp}(\bm{k},\bm{k}';i\nu_\ell)&=%-\tilde{\eta}_k\tilde{\eta}_{k'}
    V_{\rm eff}(\bm{k},\bm{k}')
    -T\sum_{\bm{q},i\omega_n}
    %\tilde{\eta}_{k}\tilde{\eta}_{q}
    V_{\rm eff}(\bm{k},\bm{q})
    G_{\rm p\uparrow}^T(\bm{q},i\omega_n+i\nu_\ell)
    G_{\rm p\downarrow}^T(-\bm{q},-i\omega_n)
    T_{\rm pp}(\bm{q},\bm{k}';i\nu_\ell).
\end{align}
The separability of the $T$-matrix $T_{\rm pp}(\bm{k},\bm{k}';i\nu_\ell)\equiv R(i\nu_\ell)\tilde{\eta}_{k}\tilde{\eta}_{k'}$
leads to
\begin{align}
    R(i\nu_\ell)&=-1+\Pi_{\rm pp}(i\nu_\ell)R(i\nu_\ell)\cr
    &=-\frac{1}{1-\Pi_{\rm pp}(i\nu_\ell)},
\end{align}
where
\begin{align}
    \Pi_{\rm pp}(i\nu_\ell)&= T\sum_{\bm{q},i\omega_n}\tilde{\eta}_{q}^2G_{\rm p\uparrow}^T(\bm{q},i\omega_n+i\nu_\ell)
    G_{\rm p\downarrow}^T(-\bm{q},-i\omega_n)\cr
    &=-\sum_{\bm{q}}\tilde{\eta}_q^2\frac{Z^2\theta(\zeta k_{\rm F}-q)}{i\nu_\ell-q^2/M_{\rm eff}-2E_{\rm P}+2\mu_{\rm p}}
\end{align}
is the two-polaronic-proton propagator.
Here we have taken the limit of $f(\xi_{\bm{q},{\rm p}})\rightarrow 0$ because $\mu_{\rm p}\rightarrow-\infty$ keeps the small proton fraction limit ($\rho_{\rm p}\rightarrow 0$).  
In this case, the problem is reduced to the effective two-body problem.
After performing the analytical continuation to the real energy (i.e., $i\nu_\ell\rightarrow \nu-2\mu_{\rm p}$), we obtain the bound-state equation for the diproton binding energy $\nu=-E_{\rm PP}+2E_{\rm P}$ (measured from the continuum bottom $2E_{\rm P}$ of two polarons) from $1-\Pi_{\rm pp}(-E_{\rm PP}+2E_{\rm P})=0$ as
\begin{align}
    1-Z^2\sum_{\bm{q}}\frac{\theta(\zeta k_{\rm F}-q)M_{\rm eff}\tilde{\eta}_q^2}{q^2+M_{\rm eff}E_{\rm PP}}=0.
\end{align}
The momentum integration reads
\begin{align}
    \sum_{\bm{q}}\frac{\theta(\zeta k_{\rm F}-q)M_{\rm eff}\tilde{\eta}_q^2}{q^2+M_{\rm eff}E_{\rm PP}}
    &=\frac{M_{\rm eff}\tilde{u}_{\rm s}^2}{2\pi^2}
    \int_0^{\zeta k_{\rm F}}q^2dq\frac{1}{(q^2+\Lambda_{\rm s}^2)^2(q^2+M_{\rm eff}E_{\rm PP})}\cr
    &=
    \frac{M_{\rm eff}\tilde{u}_{\rm s}^2}{2\pi^2}
    \left[\frac{\zeta k_{\rm F}}{2 (\Lambda_{\rm s}^2 - M_{\rm eff}E_{\rm PP}) (\Lambda_{\rm s}^2 + \zeta^2 k_{\rm F}^2)} + \frac{(M_{\rm eff}E_{\rm PP} + \Lambda_{\rm s}^2) \tan^{-1}\left(\frac{\zeta k_{\rm F}}{\Lambda_{\rm s}}\right)}{2 \Lambda_{\rm s}(\Lambda_{\rm s}^2 - M_{\rm eff}E_{\rm PP})^2}\right.\cr
    &\quad\quad\quad\quad\left.- \frac{\sqrt{M_{\rm eff}E_{\rm PP}} \tan^{-1}\left(\frac{\zeta k_{\rm F}}{\sqrt{M_{\rm eff}E_{\rm PP}}}\right)}{(\Lambda_{\rm s}^2 - M_{\rm eff}E_{\rm PP})^2}\right]. 
\end{align}
Then, the bound state equation can be rewritten as
\begin{align}
    %1&=Z^2\frac{M_{\rm eff}\Lambda_{\rm s}}{4\pi^2}
    %\left[\frac{1}{\frac{M\Lambda_{\rm s}}{8\pi}-\frac{M}{4\pi a_{\rm s}}}+\frac{10a_{\rm t}^2k_{\rm F}/M}{\left(1-\frac{a_{\rm t}\Lambda_{\rm t}}{2}\right)^2\left(1+\frac{k_{\rm F}^2}{4\Lambda_{\rm t}^2}\right)^4}\right]
    %\cr
    %&\times\left[\frac{k_{\rm F}/\Lambda_{\rm s}}{ \left(1 - \frac{M_{\rm eff}E_{\rm PP}}{\Lambda_{\rm s}^2}\right) \left(1 + \frac{k_{\rm F}^2}{\Lambda_{\rm s}^2}\right)} + \frac{\left(1+\frac{M_{\rm eff}E_{\rm PP}}{\Lambda_{\rm s}^2}\right) \tan^{-1}\left(\frac{k_{\rm F}}{\Lambda_{\rm s}}\right)
    %-2\sqrt{\frac{M_{\rm eff}E_{\rm PP}}{\Lambda_{\rm s}^2}} \tan^{-1}\left(\frac{k_{\rm F}}{\sqrt{M_{\rm eff}E_{\rm PP}}}\right)
    %}{ \left(1 - \frac{M_{\rm eff}E_{\rm PP}}{\Lambda_{\rm s}^2}\right)^2}
    %\right] \cr
    1&=\frac{2Z^2M_{\rm eff}}{\pi M}
    \left[\frac{1}{1-\frac{2}{ \Lambda_{\rm s} a_{\rm s}}}+\frac{48a_{\rm t}^2k_{\rm F}}{\pi\Lambda_{\rm s}\left(1-\frac{a_{\rm t}\Lambda_{\rm t}}{2}\right)^2\left(1+\frac{k_{\rm F}^2}{4\Lambda_{\rm t}^2}\right)^4}\right]
    \cr
    &\times\left[\frac{\zeta k_{\rm F}/\Lambda_{\rm s}}{ \left(1 - \frac{M_{\rm eff}E_{\rm PP}}{\Lambda_{\rm s}^2}\right) \left(1 + \frac{\zeta^2 k_{\rm F}^2}{\Lambda_{\rm s}^2}\right)} + \frac{\left(1+\frac{M_{\rm eff}E_{\rm PP}}{\Lambda_{\rm s}^2}\right) \tan^{-1}\left(\frac{\zeta k_{\rm F}}{\Lambda_{\rm s}}\right)
    -2\sqrt{\frac{M_{\rm eff}E_{\rm PP}}{\Lambda_{\rm s}^2}} \tan^{-1}\left(\frac{\zeta k_{\rm F}}{\sqrt{M_{\rm eff}E_{\rm PP}}}\right)
    }{ \left(1 - \frac{M_{\rm eff}E_{\rm PP}}{\Lambda_{\rm s}^2}\right)^2}
    \right]. 
\end{align}
Finally, the threshold neutron density for the presence of bound diprotons can be calculated by setting $E_{\rm pp}=0$ as
\begin{align}
    1&=\frac{2Z^2M_{\rm eff}}{\pi M}
    \left[\frac{1}{1-\frac{2}{ \Lambda_{\rm s} a_{\rm s}}}+\frac{48a_{\rm t}^2k_{\rm F}}{\pi\Lambda_{\rm s}\left(1-\frac{a_{\rm t}\Lambda_{\rm t}}{2}\right)^2\left(1+\frac{k_{\rm F}^2}{4\Lambda_{\rm t}^2}\right)^4}\right]
    \left[\frac{\zeta k_{\rm F}/\Lambda_{\rm s}}{1 + \frac{\zeta^2 k_{\rm F}^2}{\Lambda_{\rm s}^2}} + \tan^{-1}\left(\frac{\zeta k_{\rm F}}{\Lambda_{\rm s}}\right)
    \right]. 
\end{align}

\begin{figure}[t]
    \centering
    \includegraphics[width=8cm]{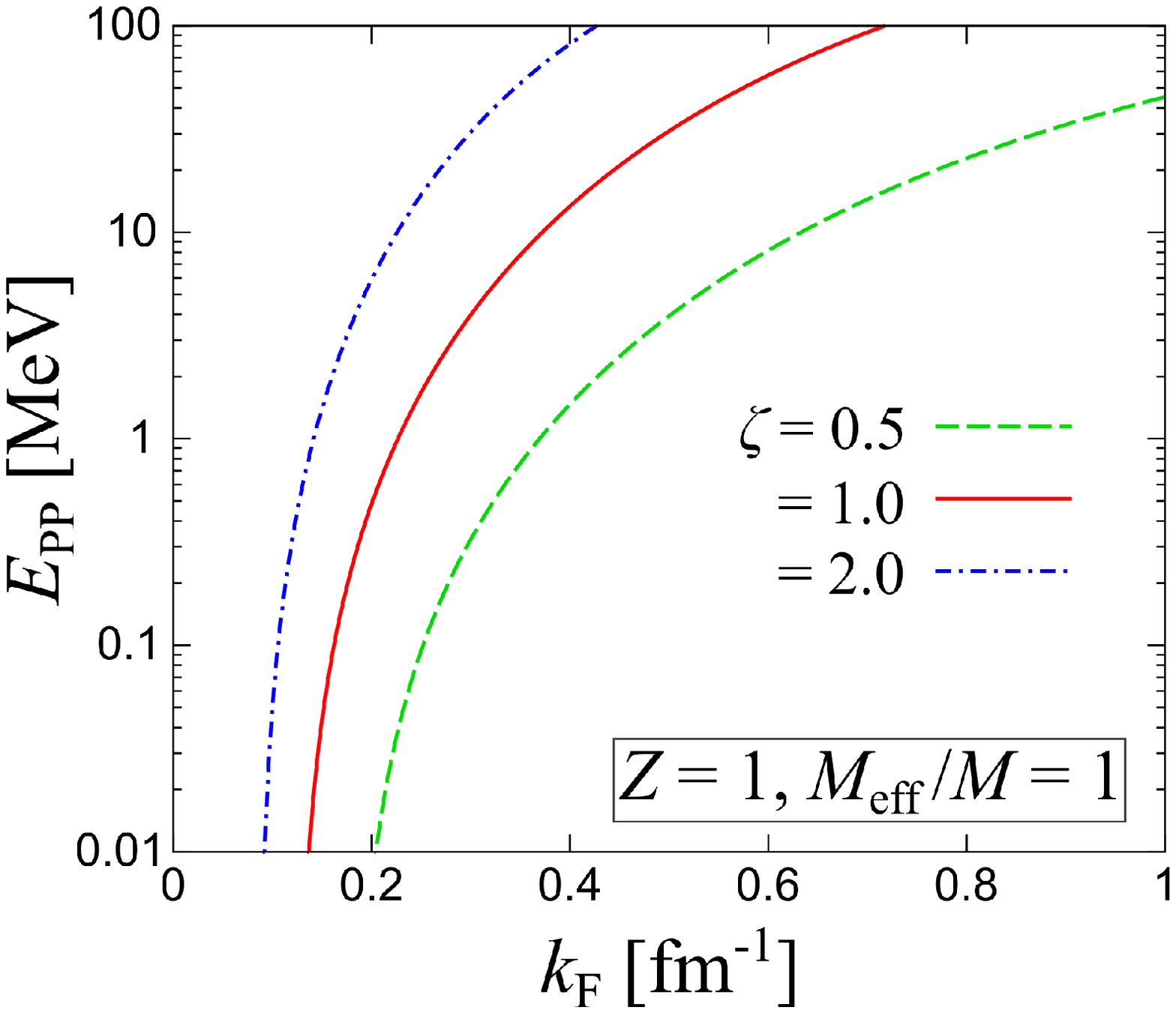}
    \caption{Cutoff dependence (i.e., $\zeta$ dependence) of the diproton binding energy $E_{\rm PP}$, as plotted for $Z=1$ and $M_{\rm eff}/M=1$.}
    \label{fig:s6}
\end{figure}
Figure~\ref{fig:s6}
shows the $\zeta$ dependence of $E_{\rm PP}$.
For simplicity, we here take $Z=1$ and $M_{\rm eff}/M=1$.
$E_{\rm PP}$ increases with $\zeta$. 
It should be noted that the large enhancement of $E_{\rm PP}$ at high neutron densities with $\zeta=2$ stems partially from an artifact associated with the low-energy approximation of $V_{\rm med.}$, which has to be weakened by the neutron Fermi degeneracy at higher momentum transfer than $k_{\rm F}$.
Although $E_{\rm PP}$ still contains uncertainties due to the $\zeta$ dependence,
the neutron density dependence of $E_{\rm PP}$ seems qualitatively robust.
A more precise calculation of $E_{\rm PP}$ is left for interesting future work.
In spite of such uncertainties, one may conclude that the threshold neutron density is located around $k_{\rm F} =O(0.1)$ fm$^{-1}$ for $\zeta=O(1)$. 

%In particular, the critical effective mass ($M_{\rm eff,c}=M_{\rm eff}$ at $E_{\rm PP}=0$) reads
%\begin{align}
%    0&=1-
%    \frac{Z^2M_{\rm eff}u^2}{2\pi^2}
%    \left[\frac{k_{\rm F}}{2 \Lambda_{\rm s}^2  (\Lambda_{\rm s}^2 + k_{\rm F}^2)} + \frac{  \tan^{-1}\left(\frac{k_{\rm F}}{\Lambda_{\rm s}}\right)}{2 \Lambda_{\rm s}^3}\right]\cr
%    &=\frac{\Lambda_{\rm s}^4}{u_{\rm s}^2}
%    -\frac{Z^2M_{\rm eff}}{4\pi^2}
%    \left[\frac{k_{\rm F}}{1 + k_{\rm F}^2/\Lambda_{\rm s}^2} +  \Lambda_{\rm s}\tan^{-1}\left(\frac{k_{\rm F}}{\Lambda_{\rm s}}\right)\right]\cr
%    &=\frac{M\Lambda_{\rm s}}{8\pi}-\frac{M}{4\pi a_{\rm s}}
%    -\frac{Z^2M_{\rm eff}}{4\pi^2}
%    \left[\frac{k_{\rm F}}{1 + k_{\rm F}^2/\Lambda_{\rm s}^2} +  \Lambda_{\rm s}\tan^{-1}\left(\frac{k_{\rm F}}{\Lambda_{\rm s}}\right)\right]\cr
%&=\frac{1}{4}-\frac{1}{2 a_{\rm s}\Lambda_{\rm s}}
%    -\frac{Z^2M_{\rm eff}}{2\pi M}
%    \left[\frac{k_{\rm F}/\Lambda_{\rm s}}{1 + k_{\rm F}^2/\Lambda_{\rm s}^2} +  \tan^{-1}\left(\frac{k_{\rm F}}{\Lambda_{\rm s}}\right)\right],    
%\end{align}
%\begin{align}
%    \frac{M}{M_{\rm eff,c}}
%    =\frac{2Z^2}{\pi}
    %\left(1-\frac{2}{a_{\rm s}\Lambda_{\rm s}}\right)
%    \left[\frac{1}{1-\frac{2}{ \Lambda_{\rm s} a_{\rm s}}}+\frac{80a_{\rm t}^2k_{\rm F}}{\pi\Lambda_{\rm s}\left(1-\frac{a_{\rm t}\Lambda_{\rm t}}{2}\right)^2\left(1+\frac{k_{\rm F}^2}{4\Lambda_{\rm t}^2}\right)^4}\right]
%    \left[\frac{k_{\rm F}/\Lambda_{\rm s}}{1 + k_{\rm F}^2/\Lambda_{\rm s}^2} +  \tan^{-1}\left(\frac{k_{\rm F}}{\Lambda_{\rm s}}\right)\right].
%\end{align}

\end{widetext}

\end{document}